\documentclass[12pt]{article}
\usepackage{feynmp,epsfig,graphics,cite}
\makeatletter
\newdimen\@bls
\@bls=\baselineskip
\newbox\slashbox \setbox\slashbox=\hbox{$/$}
\newbox\Slashbox \setbox\Slashbox=\hbox{\large$/$}
\def\section{\@startsection{section}{1}{\z@}{1.5\@bls
  \@plus .4\@bls \@minus .1\@bls}{\@bls}{\normalsize\bfseries}}
\def\subsection{\@startsection{subsection}{2}{\z@}{\@bls
  \@plus .3\@bls \@minus .1\@bls}{\@bls}{\normalsize\itshape}}
\def\subsubsection{\@startsection{subsubsection}{3}{\z@}{\@bls
  \@plus .2\@bls}{0.0001pt}{\normalsize\itshape}}
\def\paragraph{\@startsection{paragraph}{4}{\z@}{3.25ex \@plus
  2ex \@minus 0.2ex}{-1em}{\normalsize\bfseries}}
\def\pFMslash#1{\setbox\@tempboxa=\hbox{$#1$}
  \@tempdima=0.5\wd\slashbox \advance\@tempdima 0.5\wd\@tempboxa
  \copy\slashbox \kern-\@tempdima \box\@tempboxa}
\def\pFMSlash#1{\setbox\@tempboxa=\hbox{$#1$}
  \@tempdima=0.5\wd\Slashbox \advance\@tempdima 0.5\wd\@tempboxa
  \copy\Slashbox \kern-\@tempdima \box\@tempboxa}
\def\FMslash{\protect\pFMslash}

\def\half{{\textstyle\frac{1}{2}}}

\def\tr{{\rm tr} \,}

\def\m0{m^{\!\!\!\!^o}}

\def\partialslash{\FMslash \partial}

\begin{document}
$ $ \hfill GSI-Preprint-01-13 and ECT-Preprint-01-16
\begin{center}
{\Large \bf Covariant meson-baryon scattering\\ with\\[+2mm] chiral and large $N_c$
constraints\footnote{
Invited talk given by M.F.M.L. at the IARD 2000 meeting.}}\\[1cm]
{\sc M.F.M. Lutz$^a$ and E.E. Kolomeitsev$^{b}$}\\[5mm]
{\it $^a$ Gesellschaft f\"ur Schwerionenforschung (GSI),\\
Planck Str. 1, D-64291 Darmstadt, Germany}\\[3mm]
{\it $^b$ ECT$^*$, Villa Tambosi, I-38050 \,Villazzano  (Trento) \\
and INFN, G.C.\ Trento, Italy}
\end{center}
\begin{abstract}
We give a review of recent progress on the application of the
relativistic chiral $SU(3)$ Lagrangian to meson-baryon scattering.
It is shown that a combined chiral and $1/N_c$ expansion of the Bethe-Salpeter
interaction kernel leads to a good description of the kaon-nucleon, antikaon-nucleon and pion-nucleon
scattering data typically up to laboratory momenta of $p_{\rm lab} \simeq $ 500 MeV.
We solve the covariant coupled channel Bethe-Salpeter equation with the interaction kernel
truncated to chiral order $Q^3$ where we include only those terms which are
leading in the large $N_c$ limit of QCD.
\end{abstract}

\section{Introduction}

An outstanding but still thrilling problem of modern particle physics is to unravel systematically the
properties of quantum chromodynamics (QCD) in its low-energy phase where the effective degrees of
freedom are hadrons rather than quarks and gluons. There are two promising paths along which achieved one expects
significant further progress. A direct evaluation of many observable quantities is feasible by large scale
numerical simulations where QCD is put on a finite lattice. Though many static properties, like hadron
ground-state properties, have been successfully reproduced by lattice calculations, the description of
the wealth of hadronic scattering data is still outside the scope of that approach \cite{lattice:1}. That is a challenge
since many of the most exciting phenomena of QCD, like the zoo of meson and baryon resonances, are reflected
directly in the scattering data. Here a complementary approach, effective field theory, is more promising. Rather
than solving low-energy QCD directly in terms of quark and gluonic degrees of freedom, inefficient
at low energies, one aims at constructing an effective field theory in terms of hadrons directly. The idea
is to constrain that theory by as many properties of QCD as possible. That leads to a significant parameter
reduction and a predictive power of the effective field theory approach. In this spirit many effective field
theory models, not all of which fully systematic, have been constructed and applied to the data set.
Given the many empirical data points to be described, it is not always necessary to build in all
constraints of QCD. Part of QCD's properties may enter the model indirectly once it successfully
describes the data. The most difficult part in constructing an acceptable effective field theory
is the identification of its applicability domain and its accuracy level. An effective field theory,
which meets above criterium, is the so called chiral perturbation theory ($\chi$PT) applicable
in the flavor $SU(2)$ sector of low-energy QCD.

The merit of standard $\chi$PT is that first it is based on an effective Lagrangian density
constructed in accordance with all chiral constraints of QCD and second that it permits a systematic
evaluation applying formal power counting rules \cite{Weinberg}. There is mounting empirical evidence
that the QCD ground state breaks the chiral $SU(2)$ symmetry spontaneously in the limiting case where
the up and down current quark masses of the QCD Lagrangian vanish. For instance the observation that hadrons
do not occur in parity doublet states directly reflects that phenomenon. Also the smallness of the pion masses
with $m_\pi \simeq $ 140 MeV much smaller than the nucleon mass $m_N \simeq $ 940 MeV, naturally fits
into this picture once the pions are identified to be the Goldstone bosons of that spontaneously broken
chiral symmetry. In $\chi$PT the finite but small values of the up and down quark masses $m_{u,d} \simeq $ 10 MeV
are taken into account as a small perturbation defining the finite masses of the Goldstone bosons.
The smallness of the current quark masses on the typical chiral scale of $1$ GeV explains the success of
standard $\chi$PT. For early applications of $\chi$PT to pion-nucleon scattering see \cite{Gasser,Bernard,Meissner}.

It is of course tempting to generalize the successful chiral $SU(2)$ scheme to the $SU(3)$ flavor group.
To construct the appropriate chiral $SU(3)$ Lagrangian is mathematically straightforward and has been done
long ago (see e.g. \cite{Krause}). The mass $m_s \simeq 10\,m_{u,d}$ of the strange quark, though
much larger than the up and down quark masses, is still small as compared to the typical chiral scale of
1 GeV \cite{GL85}. The required approximate Goldstone boson octet is readily found with the pions, kaons and the eta-meson.
Nevertheless, important predictions of standard $\chi $PT as applied to the $SU(3)$ flavor group are in stunning
conflict with empirical observations. Most spectacular is the failure of the Weinberg-Tomozawa theorem \cite{WT}
which predicts an attractive $K^-$-proton scattering length, rather than the observed large and repulsive
scattering length.

Progress was made upon accepting the crucial observation that the power counting rules
must not be applied to a certain subset of Feynman diagrams ~\cite{weinberg:1990,lutz:1996}. Whereas for irreducible
diagrams the chiral power counting rules are well justified, this is no longer necessarily the case for the irreducible
diagrams \cite{Weinberg}. The latter diagrams are enhanced as compared to irreducible diagrams and therefore
may require a systematic resummation scheme in particular in the strangeness sectors. First intriguing works taking
up this idea in the chiral context are given in  \cite{Kaiser,Ramos,Hirschegg,Oller-Meissner}.

In this contribution we report on our recent application of the relativistic chiral $SU(3)$ Lagrangian
\cite{Hirschegg,Lutz:Kolo} to meson-baryon scattering. Preliminary results of that work were presented by one
of the authors {(M.F.M.L.) at this meeting}.  As to our knowledge this is the first application of the chiral $SU(3)$
Lagrangian density to the kaon-nucleon and antikaon-nucleon system including systematically constraints from the
pion-nucleon sector. In our scheme, which considers constraints from chiral and large $N_c$ sum rules, the baryon decuplet
field with with $J^P\!= \!\frac{3}{2}^+$ is an important ingredient, because it is part of the baryon ground state
multiplet which arises in the large $N_c$ limit of QCD \cite{Hooft,Witten}. We solve the Bethe-Salpeter equation for
the scattering amplitude with the interaction kernel truncated at chiral order $Q^3$ where we include only those terms
which are leading in the large $N_c$ limit of QCD \cite{Hooft,Witten,DJM,Carone,Luty}. The renormalization scheme is
an essential input of our chiral $SU(3)$ dynamics, because it leads to consistency with chiral counting rules and an
approximate crossing symmetry of the subthreshold scattering amplitudes. The existing low-energy cross section
data on kaon-nucleon and antikaon-nucleon scattering including angular distributions are reproduced to good accuracy.
At the same time we achieve a good description of the low-energy s- and p-wave pion-nucleon phase shifts as well as
the empirical axial-vector coupling constants of the baryon octet states.

\section{Relativistic chiral $SU(3)$ interaction terms in large $N_c$ QCD}

Details on the systematic construction principle for the chiral Lagrangian can be found, e.g., in \cite{Krause,gss,HL}.
Here we collect the interaction terms of the relativistic chiral
$SU(3)$ Lagrangian density relevant for the meson-baryon scattering
process. The basic building blocks of the chiral Lagrangian are
\begin{eqnarray}
&& U_\mu = \frac{1}{2}\,e^{-i\,\frac{\Phi}{2\,f}} \left(
\partial_\mu \,e^{i\,\frac{\Phi}{f}}
+ i\,\Big[A_\mu , e^{i\,\frac{\Phi}{f}} \Big]_+
\right) e^{-i\,\frac{\Phi}{2\,f}}  \;,\qquad \!\!
B \;, \qquad \! \!\Delta_\mu \,,
\label{def-fields}
\end{eqnarray}
where we include the pseudo-scalar meson octet field $\Phi(J^P\!\!=\!0^-)$, the baryon octet field
$B(J^P\!\!=\!{\textstyle{1\over2}}^+)$ and the baryon decuplet field
$\Delta_\mu(J^P\!\!=\!{\textstyle{3\over2}}^+)$. In (\ref{def-fields}) we consider
an external axial-vector source function $A_\mu $ which is required for the systematic
evaluation of matrix elements of the axial-vector current. A corresponding term
for the vector current is not shown in (\ref{def-fields}) because it will
not be needed in this work. The axial-vector source function
$A^\mu =\sum A_a^\mu\,\lambda^{(a)} $,
the meson octet field $\Phi=\sum \Phi_a\,\lambda^{(a)}$ and the baryon octet fields
$B= \sum B_a\,\lambda^{(a)}/\sqrt{2}$ are decomposed with the
Gell-Mann matrices $\lambda_a$ normalized with $\tr \lambda_a \,\lambda_b =
2\,\delta_{ab}$. The baryon decuplet field $\Delta^{abc} $ is completely symmetric
\begin{eqnarray}
\begin{array}{llll}
\Delta^{111} = \Delta^{++}\,, & \Delta^{113} =\Sigma^{+}/\sqrt{3}\,, &
\Delta^{133}=\Xi^0/\sqrt{3}\,,  &\Delta^{333}= \Omega^-\,, \\
\Delta^{112} =\Delta^{+}/\sqrt{3}\,, & \Delta^{123} =\Sigma^{0}/\sqrt{6}\,, &
\Delta^{233}=\Xi^-/\sqrt{3}\,, & \\
\Delta^{122} =\Delta^{0}/\sqrt{3}\,, & \Delta^{223} =\Sigma^{-}/\sqrt{3}\,, &
& \\
\Delta^{222} =\Delta^{-}\,. & & &
\end{array}
\label{dec-field}
\end{eqnarray}
The parameter $f $ in (\ref{def-fields}) can be related to the weak decay
constant of the charged pions and kaons modulo chiral $SU(3)$ correction terms. Taking
the average of the empirical decay parameters $f_\pi = 92.42 \pm 0.33 $
MeV  and $f_K \simeq 113.0 \pm 1.3$ MeV \cite{fpi:exp} one obtains the naive estimate
$f \simeq  104$ MeV. This value is still within reach of the more detailed analysis
\cite{GL85} which lead to $f_\pi/f = 1.07 \pm 0.12$. As was emphasized in \cite{MO01}
the precise value of $f$ is subject to large uncertainties.

Explicit chiral symmetry-breaking effects are included in terms
of scalar and pseudo-scalar source fields $\chi_\pm $ proportional to the quark-mass
matrix of QCD,
\begin{eqnarray}
\chi_\pm = \frac{1}{2} \left(
e^{+i\,\frac{\Phi}{2\,f}} \,\chi_0 \,e^{+i\,\frac{\Phi}{2\,f}}
\pm e^{-i\,\frac{\Phi}{2\,f}} \,\chi_0 \,e^{-i\,\frac{\Phi}{2\,f}}
\right)
\label{def-chi}
\end{eqnarray}
where $\chi_0 \sim {\rm diag} (m_u,m_d,m_s)$.
All fields in (\ref{def-fields}) and (\ref{def-chi}) have identical properties under
chiral $SU(3)$ transformations. The chiral Lagrangian consists of all possible interaction
terms, formed with the fields $U_\mu, B, \Delta_\mu $ and $\chi_\pm$ and their
respective covariant derivatives. Derivatives of the fields must be included in compliance
with the chiral $SU(3)$ symmetry. This leads to the notion of a covariant derivative
${\mathcal D}_\mu$ which is identical for all fields in (\ref{def-fields}) and (\ref{def-chi}),
\begin{eqnarray}
\Big[{\mathcal D}_\mu , B\Big]_- &=& \partial_\mu \,B +
\frac{1}{2}\,\Big[ e^{-i\,\frac{\Phi}{2\,f}} \left(
\partial_\mu \,e^{+i\,\frac{\Phi}{2\,f}}\right)
+e^{+i\,\frac{\Phi}{2\,f}} \left(
\partial_\mu \,e^{-i\,\frac{\Phi}{2\,f}}\right), B\Big]_-
\nonumber\\
&+& \frac{i}{2}\,\Big[ e^{-i\,\frac{\Phi}{2\,f}} \,
A_\mu \,e^{+i\,\frac{\Phi}{2\,f}}- e^{+i\,\frac{\Phi}{2\,f}} \, A_\mu
\,e^{-i\,\frac{\Phi}{2\,f}}, B\Big]_-  \,,
\end{eqnarray}
where we illustrated the action of  ${\mathcal D}_\mu $ at hand of the baryon octet field.

The chiral Lagrangian becomes a powerful tool once
it is combined with appropriate
power counting rules leading to a systematic approximation strategy.
One aims at describing hadronic interactions at low energy by constructing an expansion
in small momenta and the small pseudo-scalar meson masses. The infinite set
of Feynman diagrams are sorted according to their chiral powers. The minimal chiral
power $Q^{\nu }$ of a given relativistic Feynman diagram,
\begin{eqnarray}
\nu = 2-{\textstyle {1\over2}}\, E_B + 2\, L
+\sum_i V_i \left( d_i +{\textstyle {1\over2}}\, n_i-2 \right) \;,
\label{q-rule}
\end{eqnarray}
is given in terms of the number of loops, $L$, the number, $V_i$, of vertices of type $i$
with $d_i$ 'small' derivatives and $n_i$ baryon fields involved, and
the number of external baryon lines $E_B$ \cite{Weinberg}. Here one calls a derivative small
if it acts on the pseudo-scalar meson field or if it probes the virtuality of a baryon field.
Explicit chiral symmetry-breaking effects are perturbative and included in the counting scheme
with $\chi_0 \sim Q^2$. For a discussion of the relativistic chiral Lagrangian and its required
systematic regrouping of interaction terms we refer to \cite{nn-lutz,Lutz:Kolo}. The relativistic chiral
Lagrangian requires a non-standard renormalization scheme. The $MS$ or $\overline{MS}$ minimal subtraction
schemes of dimensional regularization do not comply with the chiral counting rule \cite{Gasser}. However,
an appropriately modified subtraction scheme for relativistic Feynman diagrams leads to manifest chiral
counting rules \cite{nn-lutz,Becher,Gegelia,Lutz:Kolo}. Alternatively one may work with the chiral
Lagrangian in its heavy-fermion representation \cite{J&M} where an appropriate frame-dependent
redefinition of the baryon fields lead to a more immediate  manifestation of the chiral
power counting rule (\ref{q-rule}).

In the $\pi N$ sector the $SU(2)$ chiral Lagrangian was successfully
applied \cite{Bernard,Gasser} demonstrating good convergence properties of the perturbative
chiral expansion. In the $SU(3)$ sector the situation is more involved due in part
to the relatively large kaon mass $m_K \simeq m_N/2$. Here the perturbative
evaluation of the chiral Lagrangian cannot be justified and one must change
the expansion strategy. Rather than expanding directly the scattering amplitude one may
expand the interaction kernel according to chiral power counting rules \cite{Weinberg,LePage}.
The scattering amplitude then follows from the solution of a scattering equation like the
Lipmann-Schwinger or the Bethe-Salpeter equation. This is by analogy with the treatment of the
$e^+\,e^-$ bound-state problem of QED where a perturbative evaluation of the interaction kernel
can be justified. The rational behind this change of scheme lies in the observation that
reducible diagrams are typically enhanced close to their unitarity threshold.
The enhancement factor of $(2\pi)^n$, measured relative to a reducible diagram with the
same number of independent loop integrations, is given by
the number, $n$, of reducible meson-baryon pairs in the diagram, i.e. the number of unitary
iterations implicit in the diagram. In the $\pi N$ sector this enhancement factor does not
prohibit a perturbative treatment, because the typical expansion parameter
$ m^2_\pi/(8 \pi \,f^2) \sim 0.1 $ remains sufficiently small. In the $\bar K N$ sector,
on the other hand, the factor $(2\pi)^n$ invalidates a perturbative
treatment, because the typical expansion parameter would be $m^2_K/(8 \pi\,f^2) \sim 1$.
This is in contrast to irreducible diagrams. They yield the typical expansion parameters
$m_\pi/(4 \pi \,f)$ and $m_K/(4\pi \,f)$ which justifies the perturbative
evaluation of the scattering kernels.

In our work we consider all terms of chiral order $Q^3$ in the scattering kernel
which are leading in the large $N_c$ limit. According to this philosophy loop
corrections to the Bethe-Salpeter kernel need not to be evaluated, because they carry minimal
chiral order $Q^3$ but are at the same time $1/N_c$ suppressed. In the following we will discuss all chiral
interaction terms relevant to chiral order $Q^2$. Further terms of order $Q^3$  considered in our
approach can be found in \cite{Lutz:Kolo}. Therein the interested reader may also find the
many technical developments needed to solve the Bethe-Salpeter scattering equation.

The effective chiral Lagrangian
\begin{eqnarray}
{\mathcal L} = \sum_n \,{\mathcal L}^{(n)}+\sum_n\,{\mathcal L}^{(n)}_\chi
\end{eqnarray}
falls into different classes ${\mathcal L}^{(n)}$ and ${\mathcal L}^{(n)}_\chi$.
With an upper index $n$ in ${\mathcal L}^{(n)}$ we indicate the number of fields
in the interaction vertex. The lower index $\chi $ signals terms with explicit chiral
symmetry breaking. We assume charge conjugation symmetry and parity invariance in this
work. At leading chiral order the following interaction terms are required:
\begin{eqnarray}
{\mathcal L}^{(2)} &=&
\tr \bar B \left(i\,\partialslash-\m0_{[8]}\right) \, B
+\frac{1}{4}\,\tr (\partial^\mu \,\Phi )\,
(\partial_\mu \,\Phi )
\nonumber\\
&+&\tr \bar \Delta_\mu \cdot \Big(
\left( i\,\partialslash -\m0_{[10]} \right)g^{\mu \nu }
-i\,\left( \gamma^\mu \partial^\nu + \gamma^\nu \partial^\mu\right)
+i\,\gamma^\mu\,\partialslash\,\gamma^\nu
+\,\m0_{[10]} \,\gamma^\mu\,\gamma^\nu
\Big) \, \Delta_\nu \,,
\nonumber\\
{\mathcal L}^{(3)} &=&\frac{F_{[8]}}{2\,f} \,\tr  \bar B
\,\gamma_5\,\gamma^\mu \,\Big[\left(\partial_\mu\,\Phi\right),B\Big]_-
+\frac{D_{[8]}}{2\,f} \,\tr  \bar B
\,\gamma_5\,\gamma^\mu \,\Big[\left(\partial_\mu\,\Phi\right),B\Big]_+
\nonumber\\
&-&\frac{C_{[10]}}{2\,f}\,
\tr \left\{
\Big( \bar \Delta_\mu \cdot
(\partial_\nu \,\Phi ) \Big)
\Big( g^{\mu \nu}-\half\,Z_{[10]}\, \gamma^\mu\,\gamma^\nu \Big) \,
B +\mathrm{h.c.}
\right\} \;,
\nonumber\\
{\mathcal L}^{(4)}&=& \frac{i}{8\,f^2}\,\tr\bar B\,\gamma^\mu \Big[\Big[ \Phi ,
(\partial_\mu \,\Phi) \Big]_-,B \Big]_- \,,
\label{lag-Q}
\end{eqnarray}
where we use the notations $[A,B]_\pm = A\,B\pm B\,A$ for
$SU(3)$ matrices $A$ and  $B$.
The last term in (\ref{lag-Q}) reproduces  the low-energy
theorems of meson-nucleon scattering derived first by Weinberg and Tomazawa applying
current-algebra techniques~\cite{WT}.
Note that the complete chiral interaction terms which
lead to the terms in (\ref{lag-Q}) are easily recovered by replacing
$i\,\partial_\mu \,\Phi /f \to U_\mu $. Also, a derivative acting on a baryon field
must be replaced by the covariant derivative with
$\partial_\mu \,B \to [{\mathcal D}_\mu ,B ]_- $  and
$\partial_\mu \,\Delta_\nu \to [{\mathcal D}_\mu ,\Delta_\nu ]_- $ in (\ref{lag-Q}).
With $\m0_{[8]}$ and $\m0_{[10]}$ in (\ref{lag-Q}) we denote the baryon masses
in the chiral $SU(3)$ limit. Furthermore the products of an anti-decuplet field $\bar
\Delta$ with a decuplet field $\Delta$ and an octet field $\Phi$ transform as  $SU(3)$ octets
\begin{eqnarray}
\Big(\bar \Delta \cdot \Delta \Big)^a_b &=&
\bar \Delta_{bcd}\,\Delta^{acd}\,, \qquad
\Big( \bar \Delta \cdot \Phi \Big)^a_b
=\epsilon^{kla}\,\bar \Delta_{knb}\,\Phi_l^n \,,
\nonumber\\
\Big( \Phi \cdot \Delta \Big)^a_b
&=&\epsilon_{klb}\,\Phi^l_n\,\Delta^{kna} \; ,
\label{dec-prod}
\end{eqnarray}
where $\epsilon_{abc}$ is the completely anti-symmetric pseudo-tensor.

The parameters  $F_{[8]}\simeq 0.45$
and $D_{[8]}\simeq 0.80$ are constrained by the weak decay widths of the baryon octet states
\cite{Okun}  and $C_{[10]}\simeq 1.6$ can be
estimated from the hadronic decay width of the baryon decuplet states.
The parameter $Z_{[10]}$ in (\ref{lag-Q}) may be best determined in an $SU(3)$ analysis of
meson-baryon scattering. While in the pion-nucleon sector it can be absorbed into the
quasi-local 4-point interaction terms to chiral accuracy $Q^2$ \cite{Tang}
this is no longer possible in the $SU(3)$ scheme. Our detailed analysis reveals that
the parameter $Z_{[10]}$ is relevant already at order $Q^2$ if a simultaneous chiral analysis of
the pion-nucleon and kaon-nucleon scattering processes is performed.

\subsection{Large $N_c$ counting}

We briefly recall a powerful expansion strategy which follows from QCD if the
numbers of colors ($N_c$) is considered as a large number. We present a formulation best suited
for an application to the chiral Lagrangian leading to a significant parameter reduction.
The large $N_c$ scaling of  a chiral interaction
term is easily worked out using the operator analysis proposed in \cite{Dashen}.
Interaction terms involving baryon fields are represented by matrix
elements of many-body operators in the large $N_c$ ground-state baryon
multiplet $| {\mathcal B}  \rangle $. A n-body operator is the product of
n factors formed exclusively in terms of the bilinear quark-field operators
$J_i, G_i^{(a)}$ and $T^{(a)}$ being characterized fully by their commutation
relations
\begin{eqnarray}
&&[ G^{(a)}_i\,,G^{(b)}_j]_- ={\textstyle{1\over 4}}\,i\,\delta_{ij}\,f^{ab}_{\;\;\;\,c}\,T^{(c)}
+{\textstyle{1\over 2}}\,i\,\epsilon_{ij}^{\;\;\;k}
\left({\textstyle{1\over 3}}\,\delta^{ab}\,J_{k}+d^{ab}_{\;\;\;\,c}\,G^{(c)}_k\right), \;\;
\nonumber\\
&&[ J_i\,,J_j]_- =i\,\epsilon_{ij}^{\;\;\;k}\,J_{k}\, , \quad
[ T^{(a)}\,,T^{(b)}]_- =i\,f^{ab}_{\;\;\;\,c}\,T^{(c)}\,,\quad
\nonumber\\
&& [ T^{(a)}\,,G^{(b)}_i]_- =i\,f^{ab}_{\;\;\;c}\,G^{(c)}_i \;,\quad \!
[ J_i\,,G^{(a)}_j]_- =i\,\epsilon_{ij}^{\;\;\;k}\,G^{(a)}_k \;,  \quad \!
 [ J_i\,,T^{(a)}]_- = 0\;.
\label{comm}
\end{eqnarray}
The algebra (\ref{comm}), which reflects the so-called contracted spin-flavor symmetry
of QCD, leads to a transparent derivation of the many sum rules implied by the
various infinite subclasses of QCD quark-gluon diagrams as collected at a given order in the
$1/N_c$ expansion. A convenient realization of the algebra (\ref{comm}) is obtained in
terms of non-relativistic, flavor-triplet and color $N_c$-multiplet field operators
$q$ and $q^\dagger$,
\begin{eqnarray}
&& J_i = q^\dagger \Bigg( \,\frac{\sigma_i^{(q)}}{2} \otimes 1\Bigg) \,q \,, \qquad
T^{(a)} = q^\dagger \Bigg( 1 \otimes \frac{\lambda^{(a)}}{2} \Bigg) \,q \,, \;\,
\nonumber\\
&& G^{(a)}_i = q^\dagger \Bigg(\, \frac{\sigma_i^{(q)}}{2} \otimes \frac{\lambda^{(a)}}{2} \Bigg) \,q \,.
\end{eqnarray}
If the fermionic field operators $q$ and $q^\dagger $ are assigned
anti-commutation rules the algebra (\ref{comm}) follows. The Pauli spin matrices
$\sigma^{(q)}_i$ act on the two-component spinors of the fermion fields $q, q^\dagger $
and the Gell-Mann matrices $\lambda_a$ on its flavor components. Here one needs to emphasize
that the non-relativistic quark-field operators $q$ and $q^\dagger $ must not be identified
with the quark-field operators of the QCD Lagrangian \cite{DJM,Carone,Luty}. Rather, they
constitute an effective tool to represent the operator algebra (\ref{comm}) which allows an
efficient derivation of the large $N_c$ sum rules of QCD. A systematic truncation scheme
results in the operator analysis, because a $n$-body operator is assigned the suppression factor
$N_c^{1-n}$.  The analysis is complicated by the fact that  matrix elements of $ T^{(a)}$
and $G_i^{(a)}$ may be of order $N_c$ in the baryon
ground state $|{\mathcal B} \rangle$. That implies for instance that matrix elements of
the (2$n$+1)-body operator $(T_a\,T^{(a)})^n\,T^{(c)}$ are not suppressed relative to the matrix
elements of the one-body operator $T^{(c)}$. The systematic large $N_c$ operator analysis
relies on the observation that matrix elements of the spin operator $J_i$, on the other hand,
are always of order $N_c^0$. Then a set of identities shows how to systematically represent the
infinite set of many-body operators, which one may write down at a given order in the $1/N_c$
expansion, in terms of a finite number of operators. This leads to
a convenient scheme with only a finite number of operators at given order \cite{Dashen}.
We recall typical examples of the required operator identities,
\begin{eqnarray}
&&[T_a,\, T^{(a)}]_+ -  [J_i,\,J^{(i)}]_+  ={\textstyle {1\over 6}}\,N_c\,(N_c+6)\;,
\quad [ T_{a}\,,G^{(a)}_i]_+ ={\textstyle {2\over 3}}\,(3+N_c)\,J_i \;,
\nonumber\\
&& 27\,[T_a,\, T^{(a)}]_+-12\,[G_i^{(a)},\, G_a^{(i)}]_+
= 32\,[J_i,\,J^{(i)}]_+ \;,
\nonumber\\
&& d_{abc}\,[T^{(a)},\,T^{(b)}]_+  -2\,[J_{i},\,G_c^{(i)}]_+
= -{\textstyle {1\over 3}}\,(N_c+3)\,T_c \;,
\nonumber\\
&& d^a_{\;\;bc}\,[G_a^{(i)},\,G_i^{(b)}]_+
+{\textstyle{9\over 4}}\, d_{abc}\,[T^{(a)},\,T^{(b)}]_+
= {\textstyle {10\over 3}}\,[J_{i},\,G_c^{(i)}]_+ \;,
\nonumber\\
&& d_{ab}^{\;\;\; c}\,[T^{(a)},\,G^{(b)}_i]_+ =  {\textstyle{1\over 3}}\,[J_{i},\,T^{(c)}]_+
-{\textstyle{1\over 3}}\,\epsilon_{ijk}\,f_{ab}^{\;\;\;c}\,[G^{(j)}_a\,,G^{(k)}_b]_+ \;.
\label{operator-ex}
\end{eqnarray}
For instance the first identity in (\ref{operator-ex}) shows how to avoid that
infinite tower $(T_a\,T^{(a)})^n\,T^{(c)}$ discussed above. Note that the 'parameter'
$N_c$ enters in (\ref{operator-ex}) as a mean to classify the possible realizations of
the algebra (\ref{comm}).

As a first and simple example we recall the large $N_c$ structure of the 3-point vertices.
One readily establishes two operators with parameters $g$ and $h$ at leading order
in the $1/N_c$ expansion \cite{Dashen}
\begin{eqnarray}
\langle {\mathcal B}' |\, {\mathcal L}^{(3)}\,| {\mathcal B}  \rangle  =\frac{1}{f}\,
\langle {\mathcal B}' |\, g\,G_i^{(c)}+h\,J_i\,T^{(c)}| {\mathcal B}  \rangle \,
\tr \,\lambda_c\,\nabla^{(i)}\,\Phi + {\mathcal O}\left( \frac{1}{N_c}\right) \;.
\label{3-point-vertex}
\end{eqnarray}
Further possible terms in (\ref{3-point-vertex}) are either redundant or suppressed
in the $1/N_c$ expansion. For example, the two-body operator
$ i\,f_{abc}\,G_i^{(a)} \,T^{(b)}  \sim N_c^0$ is reduced by
\begin{eqnarray}
&& i\,f_{ab}^{\;\;\;c}\,\Big[G_i^{(a)} \,,T^{(b)} \Big]_- = i\,f_{ab}^{\;\;\;c}\,i\,f^{ab}_{\;\;\;\,d}\,G_i^{(d)} =
-3\,G_i^{(c)} \;. \nonumber
\end{eqnarray}
In order to make use of the large $N_c$ result it is necessary to evaluate the matrix elements
in (\ref{3-point-vertex}) at
$N_c=3$ where one has a $\bf 56$-plet with
$| {\mathcal B}  \rangle= |B(a) ,\Delta(ijk) \rangle $. Most
economically this is achieved with the completeness identity
$1=|B\rangle \langle B|+ |\Delta\rangle \langle \Delta | $ in conjunction with
\begin{eqnarray}
&&T_c\,| B_a(\chi)\rangle = i\,f_{abc}\,| B^{(b)}(\chi)\rangle \;,\qquad
J^{(i)} \,| B_a(\chi )\rangle
= \frac{1}{2}\,\sigma^{(i)}_{\chi' \chi}| B_a(\chi')\rangle \;,
\nonumber\\
&& G^{(i)}_c\,| B_a(\chi)\rangle =
\left(\frac{1}{2}\,d_{abc}
+ \frac{1}{3}\,i\,f_{abc}\right)\,\sigma^{(i)}_{\chi'\chi}\,| B^{(b)}(\chi')\rangle
\nonumber\\
&& \qquad \qquad \quad \;
+ \frac{1}{\sqrt{2}\,2}\,\Big(\epsilon_{l}^{\;jk}\,\lambda^{(c)}_{mj}
\,\lambda_{nk}^{(a)}\Big)\,S^{(i)}_{\chi' \chi}
| \Delta^{\!(lmn)}(\chi') \rangle \;,
\label{matrix-el}
\end{eqnarray}
where $S_i\,S^\dagger_j=\delta_{ij}-\sigma_i\,\sigma_j/3$ and
$\lambda_a\,\lambda_b = {\textstyle{2 \over 3}}\,\delta_{ab}
+(i\,f_{abc}+d_{abc})\,\lambda^{(c)}$. In (\ref{matrix-el}) the baryon octet states
$| B_b(\chi)\rangle $ are labeled according to their $SU(3)$ octet index $a=1,...,8$ with
the two spin states represented by $\chi=1,2$. Similarly the decuplet states
$| \Delta_{lmn}(\chi') \rangle$ are listed with $l,m,n=1,2,3$ as defined in (\ref{dec-field}).
Note that the expressions (\ref{matrix-el}) may be
verified using the quark-model wave functions for the baryon octet and decuplet states. It is
important to realize, however, that the result (\ref{matrix-el}) is much more general than
the quark-model, because it reflects the structure of the ground-state baryons in the large
$N_c$ limit of QCD only. Matching the interaction vertices of the relativistic chiral Lagrangian
onto the static matrix elements arising in the large $N_c$ operator analysis requires a
non-relativistic reduction. It is standard to decompose the
4-component Dirac fields $B$ and $\Delta_\mu $ into baryon octet and
decuplet spinor fields $B(\chi)$ and $\Delta (\chi)$:
\begin{eqnarray}
\Big(B, \Delta_\mu \Big) \to  \left(
\begin{array}{c}
 \left(\frac{1}{2}+\frac{1}{2}\,\sqrt{1+\frac{\nabla^2}{M^2}}
\right)^{\frac{1}{2}} \Big(B(\chi ),S_\mu\,\Delta (\chi )\Big) \\
\frac{(\sigma \cdot \nabla )}{\sqrt{2}\,M}
\left(1+\sqrt{1+\frac{\nabla^2}{M^2} }\,\right)^{-\frac{1}{2}}
\Big(B(\chi ),S_\mu\,\Delta (\chi )\Big)
\end{array}
\right)
\end{eqnarray}
where $M$ denotes the baryon octet and decuplet mass in the large $N_c$ limit. At
leading order one finds $S_\mu =(0, S_i) $ with the transition matrices $S_i$ introduced in
(\ref{matrix-el}). It is then straightforward to expand in powers of $\nabla/M$ and
achieve the desired matching. This leads for example to the identification $D_{[8]}=g $,
$F_{[8]}= 2\,g/3+h$ and $C_{[10]}=2\,g$. The empirical values of $F_{[8]},D_{[8]}$ and $C_{[10]}$ are
quite consistent with those relations \cite{Jenkins}. Note that operators at subleading order in
(\ref{3-point-vertex}) then parameterize the deviation from $C\simeq 2 \,D$.

\subsection{Quasi-local interaction terms}

We turn to the two-body interaction terms at chiral order $Q^2$. From phase space consideration it is evident
that at this order there are only terms which contribute to the meson-baryon s-wave scattering lengths, the
s-wave effective range parameters and the p-wave scattering volumes. Higher partial
waves are not involved at this order. The various contributions are distributed with:
\begin{eqnarray}
{\mathcal L}^{(4)}_2= {\mathcal L}^{(S)}+{\mathcal L}^{(V)}+{\mathcal L}^{(T)}
\,,
\label{l42}
\end{eqnarray}
where the lower index k in ${\mathcal L}^{(n)}_k$ denotes the minimal chiral order
of the interaction vertex. In the relativistic framework one observes mixing of the
partial waves in the sense that for instance ${\mathcal L}^{(S)}, {\mathcal L}^{(V)}$
contribute to the s-wave channels and ${\mathcal L}^{(S)}, {\mathcal L}^{(T)}$ to the
p-wave channels. We write:
\begin{eqnarray}
{\mathcal L}^{(S)}&=&\frac{g^{(S)}_0}{8\!\,f^2}\,\tr\bar B\,B
\,\tr (\partial_\mu\Phi) \, (\partial^\mu\Phi)
+\frac{g^{(S)}_1}{8\!\,f^2}\,\tr \bar B \,(\partial_\mu\Phi)
\,\tr(\partial^\mu\Phi) \, B
\nonumber\\
&+&\frac{g^{(S)}_F}{16\!\,f^2} \,\tr\bar B \Big[
\Big[(\partial_\mu\Phi),(\partial^\mu\Phi)
\Big]_+ ,B\Big]_-
\!+\frac{g^{(S)}_D}{16\!\,f^2}\,\tr \bar B \Big[
\Big[(\partial_\mu\Phi),(\partial^\mu\Phi)
\Big]_+, B \Big]_+ \;,
\nonumber\\
{\mathcal L}^{(V)}&=&
\frac{g^{(V)}_0}{16\!\, f^2}\,
\Big(\tr \bar B \,i\,\gamma^\mu\,( \partial^\nu B) \,
\tr(\partial_\nu\Phi) \, ( \partial_\mu\Phi)
+\mathrm{h.c.}\Big)
\nonumber\\
&+&\frac{g^{(V)}_1}{32\!\,f^2}\,
\tr \bar B \,i\,\gamma^\mu\,\Big( ( \partial_\mu\Phi)
\,\tr(\partial_\nu\Phi) \, ( \partial^\nu B)
+ ( \partial_\nu\Phi) \,
\tr ( \partial_\mu \Phi) \, ( \partial^\nu B)
+\mathrm{h.c.}\Big)
\nonumber\\
&+&\frac{g_F^{(V)}}{32\!\,f^2}\,\Big(
\tr   \bar B \,i\,\gamma^\mu\,\Big[
\Big[(\partial_\mu \Phi) , (\partial_\nu\Phi)\Big]_+,
( \partial^\nu B) \Big]_-
+\mathrm{h.c.} \Big)
\nonumber\\
&+& \frac{g^{(V)}_D}{32\!\,f^2}\,\Big(\tr
\bar B \,i\,\gamma^\mu\,\Big[
\Big[( \partial_\mu\Phi) , (\partial_\nu\Phi)\Big]_+,
( \partial^\nu B) \Big]_+
+\mathrm{h.c.} \Big)\,,
\nonumber\\
{\mathcal L}^{(T)}&=&
\frac{g^{(T)}_1}{8\!\,f^2}\,\tr\bar B \,( \partial_\mu\Phi)
\,i\,\sigma^{\mu \nu}\,\tr( \partial_\nu \Phi) \, B
\nonumber\\
&+& \frac{g^{(T)}_D}{16\!\,f^2}\,
\tr \bar B \,i\,\sigma^{\mu \nu}\,\Big[
\Big[(\partial_\mu \Phi) ,( \partial_\nu \Phi)  \Big]_-, B \Big]_+
\nonumber\\
&+&\frac{g^{(T)}_F}{16\!\,f^2}\,
\tr \bar B \,i\,\sigma^{\mu \nu}\,\Big[
\Big[( \partial_\mu \Phi) ,( \partial_\nu\Phi) \Big]_- ,B\Big]_-
\,.
\label{two-body}
\end{eqnarray}
It is clear that if the heavy-baryon expansion is applied to (\ref{two-body})
the quasi-local 4-point interactions can be mapped onto corresponding terms
of  the heavy-baryon formalism presented for example in \cite{CH-Lee}.

We apply the large $N_c$ counting rules in order to estimate the
relative importance of the  quasi-local $Q^2$-terms in (\ref{two-body}).
Terms which involve a single-flavor trace are enhanced as compared to the
double-flavor trace terms. This is obvious, because a flavor trace in an interaction term
is necessarily accompanied by a corresponding color trace if visualized in terms of quark
and gluon lines. A color trace signals a quark loop and therefore provides the announced
$1/N_c$ suppression factor \cite{Hooft,Witten}. The counting rules are nevertheless
subtle, because a certain combination of double trace expressions can be rewritten in
terms of a single-flavor trace term \cite{Fearing}
\begin{eqnarray}
&&\tr \left( \bar B \, B \right)
\,\tr \Big(\Phi \, \Phi  \Big)
+2\,\tr \left( \bar B \, \Phi \right)
\,\tr \Big(\Phi \, B  \Big)
\nonumber\\
=&&\tr \Big[\bar B, \Phi \Big]_-\,\Big[B, \Phi
\Big]_-
+\frac{3}{2}\,\tr \bar B \Big[
\Big[\Phi , \Phi  \Big]_+, B \Big]_+ \;.
\label{trace-id}
\end{eqnarray}
Thus one expects for example that both parameters $g_{0}^{(S)}$ and $g_{1}^{(S)}$ may be large
separately but the combination $2\,g_0^{(S)}-g_1^{(S)}$ should be small. A more detailed
operator analysis  leads to
\begin{eqnarray}
&&\langle {\mathcal B}' | {\mathcal L}^{(4)}_2 | {\mathcal B} \rangle  = \frac{1}{16\,f^2}\,
\langle {\mathcal B}' | \,O_{ab}(g_1,g_2)\, | {\mathcal B} \rangle
\,\tr [(\partial_\mu \Phi),\lambda^{(a)}]_-\,[(\partial^\mu \Phi),\lambda^{(b)}]_-
\nonumber\\
&&  \qquad \qquad +\frac{1}{16\,f^2}\,
\langle {\mathcal B}' | \,O_{ab}(g_3,g_4)\, | {\mathcal B} \rangle
\,\tr [(\partial_0 \Phi),\lambda^{(a)}]_-\,[(\partial_0 \Phi),\lambda^{(b)}]_-
\nonumber\\
&& \qquad \qquad  + \frac{1}{16\,f^2}\,
\langle {\mathcal B}' | \,O^{(ij)}_{ab}(g_5,g_6)\, | {\mathcal B} \rangle
\,\tr [(\nabla_i \Phi),\lambda^{(a)}]_-\,[(\nabla_j \Phi),\lambda^{(b)}]_- \;,
\nonumber\\ \nonumber\\
&& O_{ab}(g,h) = g\,d_{abc}\,T^{(c)} + h\,[T_a,T_b]_+
+{\mathcal O}\left(\frac{1}{N_c} \right) \;,
\nonumber\\
&& O_{ab}^{(ij)}(g,h) = i\,\epsilon^{ijk }\,i\,f_{abc}\left(
g\,G_k^{(c)} + h\,J_{k}\,T^{(c)} \right)
+{\mathcal O}\left(\frac{1}{N_c} \right) \;.
\label{Q^2-large-Nc}
\end{eqnarray}
We checked that other forms for the coupling of the operators $O_{ab}$ to the
meson fields do not lead to new structures.
It is straightforward to match the coupling constants $g_{1,..,6}$ onto the
ones in (\ref{two-body}). Identifying the leading terms in the non-relativistic
expansion we obtain:
\begin{eqnarray}
&& g_0^{(S)}= \frac{1}{2}\,g_1^{(S)} = \frac{2}{3}\,g_D^{(S)}
= -2\,g_2\,, \qquad  \; g_F^{(S)}= -3\,g_1 \,,
\nonumber\\
&& g_0^{(V)}= \frac{1}{2}\,g_1^{(V)}=\frac{2}{3}\,g_D^{(V)}
=-2\,\frac{g_4}{M}\,, \qquad
g_F^{(V)}= -3\,\frac{g_3}{M} \,,
\nonumber\\
&& g_1^{(T)}= 0 \,, \qquad
g_F^{(T)}= -g_5-\frac{3}{2}\,g_6 \,,  \qquad g_D^{(T)}= -\frac{3}{2}\,g_5 \;,
\label{Q^2-large-Nc-result}
\end{eqnarray}
where $M$ is the large $N_c$ value of the baryon octet mass.
We conclude that at chiral order $Q^2$ there are only six leading large $N_c$
coupling constants.

Part of the predictive power of the chiral Lagrangian results, because
chiral $SU(3)$ symmetry selects certain subsets of all $SU(3)$ symmetric
tensors at a given chiral order.

\subsection{Explicit chiral symmetry breaking}

There remain the interaction terms proportional to $\chi_\pm $
which break the chiral $SU(3)$ symmetry explicitly.
We collect here all relevant terms of chiral order $Q^2$ \cite{Gasser,Kaiser}.
It is convenient to visualize the symmetry-breaking
fields $\chi_\pm$ of (\ref{def-chi}) in their expanded forms:
\begin{eqnarray}
\! \!\! \chi_+ = \chi_0 -\frac{1}{8\,f^2}
\Big[ \Phi, \Big[ \Phi ,\chi_0 \Big]_+\Big]_+ \!+{\mathcal O} \left(\Phi^4 \right) \,,
\;\,
\chi_- = \frac{i}{2\,f}\, \Big[ \Phi ,\chi_0 \Big]_+
\!+{\mathcal O} \left(\Phi^3 \right) \,.
\label{chi-exp}
\end{eqnarray}
We begin with the 2-point interaction vertices which all result exclusively from chiral
interaction terms linear in $\chi_+$. They read
\begin{eqnarray}
{\mathcal L }_{\chi}^{(2)}&=& -\frac{1}{4}\,\tr \Phi\,\Big[ \chi_0, \Phi \Big]_+
+2\,\tr \bar B \left( b_D\,\Big[ \chi_0 , B \Big]_+ +b_F\,\Big[ \chi_0 , B \Big]_-
+b_0 \, B \,\tr \chi_0 \right)
\nonumber\\
&+&2\,d_D\,\tr \Big(\bar \Delta_\mu\cdot \Delta^\mu  \Big)  \,\chi_0
+2\,d_0 \,\tr \left(\bar \Delta_\mu\cdot  \Delta^\mu\right) \,\tr \chi_0
\nonumber\\
&+&\tr \bar B \left(i\,\partialslash-\m0_{[8]}\right) \,
\Big( \zeta_0\,B \, \tr  \chi_0
+  \zeta_D\,[B, \chi_0 ]_+ + \zeta_F\, [B, \chi_0 ]_- \Big) \;,
\nonumber\\ \nonumber\\
\chi_0 &=&\frac{1}{3} \left( m_\pi^2+2\,m_K^2 \right)\,1
+\frac{2}{\sqrt{3}}\,\left(m_\pi^2-m_K^2\right) \lambda_8 \;,
\label{chi-sb}
\end{eqnarray}
where we normalized $\chi_0$ to give the pseudo-scalar mesons their isospin averaged
masses. The first term in (\ref{chi-sb}) leads to the finite masses of the pseudo-scalar
mesons. Note that to chiral order $Q^2$ one has $m_\eta^2 = 4\,(m_K^2-m_\pi^2)/3$.
The parameters $b_D$, $b_F$, and $d_D$ are determined to leading order by the
baryon octet and decuplet mass splitting
\begin{eqnarray} \nonumber
&&m_{[8]}^{(\Sigma )}-m_{[8]}^{(\Lambda )}=
{\textstyle{16\over 3}}\,b_D\,(m_K^2-m_\pi^2)\,, \quad
m_{[8]}^{(\Xi )}-m_{[10]}^{(N )} =-8\,b_F\,(m_K^2-m_\pi^2)\,,
\\ \nonumber
&&m_{[10]}^{(\Sigma )}-m_{[10]}^{(\Delta )}=m_{[8]}^{(\Xi )}-m_{[10]}^{(\Sigma )}
=m_{[10]}^{(\Omega )}-m_{[10]}^{(\Xi )}
=-\textstyle{4\over 3}\, d_D\, (m_K^2-m_\pi^2)\,.
\label{mass-splitting}
\end{eqnarray}
The empirical baryon masses lead to the estimates $b_D \simeq 0.06$~GeV$^{-1}$,
$b_F \simeq -0.21$~GeV$^{-1}$, and $d_D\simeq -0.49$~GeV$^{-1}$.
For completeness we recall the leading large $N_c$ operators for
the baryon mass splitting (see e.g. \cite{Jenkins})
\begin{eqnarray}
&&\langle {\mathcal B}' |{\mathcal L }_{\chi}^{(2)} | {\mathcal B} \rangle
=\langle {\mathcal B}' |\,
b_1\,T^{(8)} + b_2\,[J^{(i)}, G^{(8)}_i]_+ \,| {\mathcal B} \rangle
+{\mathcal O}\left(\frac{1}{N_c^{2}}\right)\,,
\nonumber\\
&&b_D = -\frac{\sqrt{3}}{16}\,\frac{3\,b_2}{m_K^2-m_\pi^2} \,, \quad
b_F =-\frac{\sqrt{3}}{16}\,\frac{2\,b_1+b_2}{m_K^2-m_\pi^2}\,,\quad
d_D = -\frac{3\,\sqrt{3}}{8}\,\frac{b_1+2\,b_2}{m_K^2-m_\pi^2} \,,
\end{eqnarray}
where we matched the symmetry-breaking parts with $\lambda_8$.
One observes that the empirical values for $b_D+b_F$ and $d_D$ are remarkably consistent with
the large $N_c$ sum rule $b_D+b_F\simeq {\textstyle{1\over 3}}\,d_D$. The parameters $b_0$
and $d_0$ are more difficult to access. They determine the deviation of the octet and
decuplet baryon masses from their chiral $SU(3)$ limit values $\m0_{[8]}$ and $\m0_{[10]}$
\begin{eqnarray}
&& m_{[8]}^{(N)} = \m0_{[8]}-2\,m_\pi^2\,(b_0+2\,b_F)-4\,m_K^2\,(b_0+b_D-b_F) \;,
\nonumber\\
&& m_{[10]}^{(\Delta )}  = \m0_{[10]}-2\,m_\pi^2 (d_0 +d_D)-4\,m_K^2 \,d_0 \;,
\label{piN-sig-term}
\end{eqnarray}
where terms of chiral order $Q^3$ are neglected. The size of the parameter $b_0$ is
commonly encoded into the pion-nucleon sigma term
$$\sigma_{\pi N}= -2\,m_\pi^2\,(b_D+b_F+2\,b_0) +{\mathcal O}\left(Q^3\right)\,.$$
Note that the former standard value $\sigma_{\pi N}=(45\pm 8)$~MeV of
\cite{piN-sigterm} is currently under debate \cite{pin-news}.

The predictive power of the chiral Lagrangian lies in part in the strong correlation of
vertices of different degrees as implied by the non-linear fields $U_\mu $ and $\chi_\pm $.
A powerful example is given by the two-point vertices introduced in (\ref{chi-sb}). Since
they result from chiral interaction terms linear in the $\chi_+$-field (see (\ref{chi-exp}))
they induce particular meson-octet baryon-octet interaction vertices
\begin{eqnarray}
{\mathcal L }_{\chi}^{(4)}&=& - \frac{1}{4\,f^2}\,\tr \bar B \left(b_D\,
\Big[ \Big[ \Phi, \Big[ \Phi ,\chi_0 \Big]_+\Big]_+ , B \Big]_+
+b_F\,\Big[ \Big[ \Phi, \Big[ \Phi ,\chi_0 \Big]_+\Big]_+ , B \Big]_-
\right)
\nonumber\\
&-&\frac{b_0}{4\,f^2}\,\tr\bar{B}\, B \,\tr \Big[ \Phi, \Big[ \Phi ,\chi_0 \Big]_+\Big]_+
\;.
\label{chi-sb-4}
\end{eqnarray}
At chiral order $Q^2$ there are no further four-point interaction terms with explicit
chiral symmetry breaking.


Closing this section we would like to emphasize that one
should not be discouraged by the many fundamental parameters
introduced in this section. The empirical data set includes many hundreds of
data points and will be reproduced rather accurately. Our scheme makes many predictions
for poorly known or not known observable quantities like for example the p-wave scattering volumes
of the kaon-nucleon scattering processes or the many $SU(3)$ reactions like
$\pi \Lambda \to \pi \Sigma $. Also one should realize that in a more conventional
meson-exchange approach, besides lacking a systematic approximation scheme, many parameters
are implicit in the so-called form factors. In a certain sense the parameters used in the
form factors reflect the more systematically constructed and controlled quasi-local
counter terms of the chiral Lagrangian.

\section{Meson-baryon scattering}

We present a selection of results obtained in our detailed chiral
$SU(3)$ analysis of the low-energy meson-baryon scattering data. Before delving into details we briefly summarize
the main features and crucial arguments of our approach. We consider the number of colors ($N_c$) in QCD as a
large parameter relying on a systematic expansion of the interaction kernel in powers of $1/N_c$. The coupled-channel
Bethe-Salpeter kernel is evaluated in a combined chiral and $1/N_c$ expansion to chiral order $Q^3$. The scattering
amplitudes for the meson-baryon scattering processes are obtained from the solution of the coupled-channel Bethe-Salpeter
scattering equation. Approximate crossing symmetry of the amplitudes is guaranteed by a renormalization program which
leads to the matching of subthreshold amplitudes. After giving a discussion of baryon-resonance generation
in the coupled-channel approach, we present the parameters as they are adjusted to the data set. In the subsequent
sections we confront our results with the empirical data. Our theory is referred to as the "$\chi $-BS(3)" approach
for chiral Bethe-Salpeter dynamics of the flavor $SU(3)$ symmetry.

\subsection{Dynamical generation of baryon resonances}

\begin{figure}[t]
\begin{center}
\includegraphics[width=15cm,clip=true]{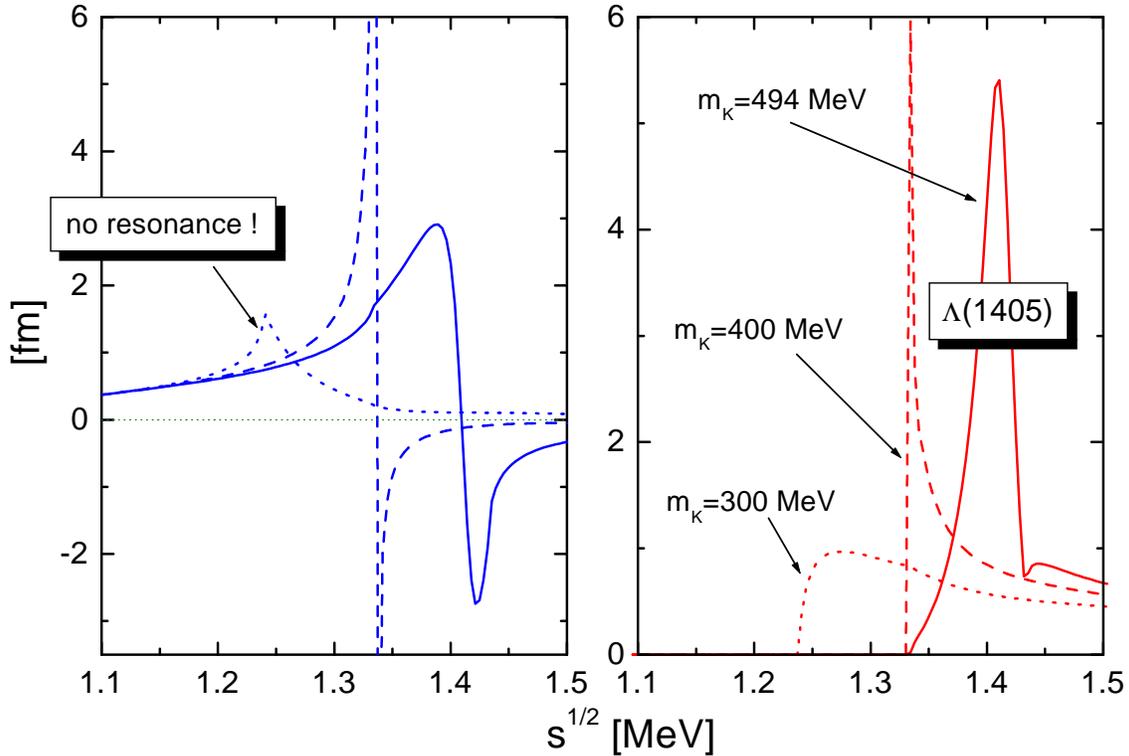}
\end{center}
\caption{Real (l.h.s.) and imaginary (r.h.s.) part of the isospin
zero s-wave $K^-$-nucleon scattering amplitude as it follows from
the $SU(3)$ Weinberg-Tomozawa interaction term in a
coupled-channel calculation. We use here $f = 93 $ MeV.}
\label{fig:wt}
\end{figure}

A qualitative understanding of the typical strength in the various channels can
be obtained already at leading chiral order $Q$. In particular the
$\Lambda(1405)$ resonance is formed as a result of the coupled-channel
dynamics defined by the leading order Weinberg-Tomozawa interaction vertices ($\mathcal{L}^{(4)}$ in
(\ref{lag-Q})). If taken as input for the multi-channel Bethe-Salpeter equation a rich structure of the
scattering amplitude arises. Fig.~\ref{fig:wt} shows the s-wave solution of the multi-channel
Bethe-Salpeter as a function of the kaon mass. For physical kaon masses the isospin zero scattering amplitude
exhibits a resonance structure at energies where one would expect the $\Lambda(1405)$
resonance. We point out that the resonance structure disappears as the kaon mass is decreased.
Already at a hypothetical kaon mass of $300$ MeV the $\Lambda(1405)$ resonance is not
formed anymore. Fig.~\ref{fig:wt} nicely demonstrates that the chiral $SU(3)$ Lagrangian is
necessarily non-perturbative in the
strangeness sector \cite{Kaiser,Ramos,Hirschegg,dalitz-1,Siegel,krippa}.
Unitary (reducible) loop diagrams are typically enhanced by a factor of $2 \pi $ close to threshold relatively
to irreducible diagrams and must therefore be summed to infinite orders.

An immediate question arises: what is the generic nature of the nucleon and hyperon resonances?
We successfully generated the s-wave resonance $\Lambda (1405)$ in terms of chiral $SU(3)$ coupled-channel
dynamics. Are more resonances of dynamic origin? We expect all baryon resonances, with the important exception
of those resonances which belong to the large $N_c$ baryon ground states, to be generated by coupled-channel dynamics.
This conjecture is based on the observation that unitary (reducible) loop diagrams are typically enhanced
by a factor of $2 \pi $ close to threshold relatively to irreducible diagrams.
That factor invalidates the perturbative evaluation  of the scattering amplitudes and leads necessarily to a
non-perturbative scheme with reducible diagrams summed to all orders.

A more conservative approach would be to explicitly incorporate additional resonance fields in the chiral Lagrangian.
That causes, however, severe problems as there is no longer a straightforward systematic approximation strategy.
Given that the baryon octet and decuplet states are degenerate in the large $N_c$ limit, it is natural to impose
$m_{[10]} -m_{[8]} \sim Q$. In contrast with that there is no fundamental reason to insist on $m_{N^*}-m_N \sim Q$.
But, only with $m_{N^*}-m_{[8]} \sim Q $ is it feasible to establish consistent power counting rules needed for the
systematic evaluation of the chiral Lagrangian. Note that the presence of the nucleon resonance states in the
large $N_c$ limit of QCD is far from obvious. Since reducible diagrams are typically enhanced by a
factor of $2 \pi$ relatively to irreducible diagrams, there are a priori two possibilities: the baryon resonances
are a consequence of important coupled-channel dynamics or they are already present in the interaction kernel.
Our opinion differs here from the one expressed in \cite{Carone-1,Carone-2} where for instance the d-wave baryon
resonance states are considered as part of an excited large $N_c$ ${\bf 70}$-plet.  An expansion in $2 \pi/N_c $, in
our world with $N_c=3$, does not appear useful. Also the fact that baryon resonances exhibit large hadronic decay
widths may be taken as an indication that the coupled-channel dynamics is the driving mechanism for the creation of
baryon resonances. Related arguments have been put forward in \cite{Aaron:1,Aaron:2}. Indeed, for instance the
d-wave $N(1520)$ resonance, was successfully described in terms of coupled-channel dynamics, including the
vector-meson nucleon channels as an important ingredient \cite{QM-lutz}, but without assuming a preformed resonance
structure in the interaction kernel. The successful description of the $\Lambda (1405)$ resonance in our scheme
(see Fig. \ref{fig:wt}) supports the above arguments. For a recent discussion of the competing picture in which
the $\Lambda (1405)$ resonance is considered as a quark-model state we refer to \cite{Kimura}.

The description of resonances has a subtle consequence for the treatment of the u-channel baryon resonance
exchange contribution. If a resonance is formed primarily by the coupled-channel dynamics one should not include
an explicit bare u-channel resonance contribution in the interaction kernel. The necessarily strong energy
dependence of the resonance self-energy would invalidate the use of the bare u-channel resonance contribution,
because for physical energies $\sqrt{s}> m_N+m_\pi$ the resonance self-energy is probed far off the resonance pole.
Our discussion has non-trivial implications for the chiral SU(3) dynamics. Naively one may object that the effects of
the u-channel baryon resonance exchange contributions can be absorbed to good accuracy into chiral two-body
interaction terms in any case. However, while this is true in a chiral $SU(2)$ scheme, this is no longer possible
in a chiral SU(3) approach. This follows because chiral symmetry selects a specific subset of all possible
SU(3)-symmetric two-body interaction terms to given chiral order. For that reason we discriminate two possibilities.
In scenario I we conjecture that the baryon resonance states are primarily generated by the
coupled-channel dynamics. This is analogous to the treatment of the $\Lambda (1405)$ resonance, which is generated
dynamically in the chiral SU(3) scheme (see Fig. \ref{fig:wt}). Here a s-channel pole term is generated by the
coupled-channel dynamics but the associated u-channel term is effectively left out as a much suppressed contribution.
In scenario II we explicitly include the s- and the u-channel resonance exchange contributions, thereby assuming
that the resonance was preformed already in the large $N_c$ limit of QCD and only slightly affected by the
coupled-channel dynamics. In a scheme based on scenario I, which however does not consider important channels
required to generate a specific resonance, we suggest to include only the s-channel resonance contribution
as a reminiscence of the neglected channels. An example is the d-wave $J^P={\textstyle{3\over 2}}^-$ baryon nonet
resonance which we believe to be generated primarily by the vector-meson baryon channels \cite{Aaron:1,Aaron:2,QM-lutz}.

Our detailed analyses of the data set clearly favor scenario I. The inclusion of a u-channel baryon-nonet d-wave
resonance exchange contributions appears to destroy the subtle balance of chiral s-wave range terms and makes it
impossible to obtain a reasonable fit to the data set. Thus all results reviewed in this work are based on scenario I.
In our present scheme we include contributions of s- and u-channel baryon
octet and decuplet states explicitly but only the s-channel contributions of an explicit
d-wave $J^P={\textstyle{3\over 2}}^-$ baryon nonet resonance states. Thereby we consider the s-channel
baryon nonet contributions to the interaction kernel as a reminiscence of additional inelastic channels not
included in the present scheme like for example the $K\,\Delta_\mu $ or $K_\mu \,N$ channel.
It is gratifying to find precursor effects for the s-wave $\Sigma (1750)$ and
the p-wave $\Lambda (1600)$, $\Lambda (1890)$ and p-wave $\Sigma (1660)$ resonances
once agreement with the low-energy data set with $p_{\rm lab.}< 500$ MeV was achieved.
A more accurate description of the latter resonances requires the extension of the $\chi$-BS(3)
approach including more inelastic channels. These finding strongly support the conjecture that all
baryon resonances but the decuplet ground states are a consequence of coupled-channels dynamics.

\subsection{Parameters}

\tabcolsep=1.4mm
\renewcommand{\arraystretch}{1.5}
\begin{table}[t]\begin{center}
\begin{tabular}{|c|c|c|c|c|}
\hline $f$ [MeV] & $ C_R$ & $F_R$ & $D_R$
\\
\hline \hline 90.04  & 1.734 & 0.418 & 0.748
\\  \hline
\end{tabular}
\end{center}
\caption{Leading chiral parameters which contribute to
meson-baryon scattering at order $Q$.} \label{q1param:tab}
\end{table}

\begin{table}[t]\begin{center}
\begin{tabular}{|c|c||c|c||c|c||c|c|}
\hline $g_F^{(V)}$[GeV$^{-2}$] & 0.293 & $g_F^{(S)}$[GeV$^{-1}$] &
-0.198 & $g_F^{(T)}$[GeV$^{-1}$] & 1.106 & $Z_{[10]}$  &  0.719
\\
\hline $g_D^{(V)}$[GeV$^{-2}$] & 1.240 & $g_D^{(S)}$[GeV$^{-1}$] &
-0.853 & $g_D^{(T)}$[GeV$^{-1}$] & 1.607 & - & -
\\
\hline
\end{tabular}
\end{center}
\caption{Chiral $Q^2$-parameters resulting from a fit to
low-energy meson-baryon scattering data. Further parameters at
this order are determined by the large $N_c$ sum rules.}
\label{q2param:tab}
\end{table}

The set of parameters is well determined by the available scattering data and weak decay widths of the
baryon octet states. We aimed at a uniform quality of the data description.
In Tabs. \ref{q1param:tab} and \ref{q2param:tab} we present the parameter set to chiral order $Q^2$
of our best fit to the data collection. Note that part of the parameters are predetermined
to a large extent and therefore fine-tuned only in a small interval.

A qualitative understanding of the typical strength in the various channels can
be obtained already at leading chiral order $Q$. In particular, as shown in Fig. ~\ref{fig:wt}, the
$\Lambda(1405)$ resonance is formed as a result of the coupled-channel
dynamics defined by the Weinberg-Tomozawa interaction vertices ($\mathcal{L}^{(4)}$ in
(\ref{lag-Q})). There are four parameters relevant at that order $f$, $C_R$, $F_R$ and $D_R$.
Their respective values as given in Tab. \ref{q1param:tab} are the result
of our global fit to the data set including all parameters of the $\chi$-BS(3) approach.
At leading chiral order the parameter $f$ determines the weak pion- and kaon-decay processes and at the same
time the strength of the Weinberg-Tomozawa interaction vertices. At subleading order the
Weinberg-Tomozawa terms and the weak-decay constants of the pseudo-scalar meson octet receive independent
correction terms. The result $f\simeq  90$ MeV is sufficiently close to the
empirical pion and kaon weak-decay constants $f_\pi \simeq 92.4$ MeV and
$f_K \simeq 113.0$ MeV to expect that the chiral correction terms lead indeed
to values rather close to the empirical decay constants. Our value for $f$ is consistent
with the estimate of \cite{GL85} which lead to $f_\pi/f = 1.07 \pm 0.12$.
The baryon octet and decuplet s- and u-channel exchange contributions to the interaction kernels are determined
by the $F_R, D_R$ and $C_R$ parameters at leading order. Note that $F_R$ and $D_R$ predict
the baryon octet weak-decay processes and $C_R$ the strong decay widths of the baryon decuplet states also
at this order.

A quantitative description of the data set requires the inclusion
of higher order terms. Initially we tried to establish a consistent
picture of the existing low-energy meson-baryon scattering data
based on a truncation of the interaction kernels to chiral order
$Q^2$. This attempt failed due to the insufficient quality of the
kaon-nucleon scattering data at low energies. In particular some of the
inelastic $K^-$-proton differential cross sections are strongly influenced by
the d-wave $\Lambda(1520)$ resonance at energies where the data points
start to show smaller error bars.  We conclude that, on the one hand,
one must include an effective baryon-nonet resonance field and, on the other
hand, perform minimally a chiral $Q^3$ analysis to extend the
applicability domain to somewhat higher energies.
Since the effect of the d-wave resonances is only necessary in the strangeness minus one sector,
they are only considered in that channel.

At subleading order $Q^2$ the chiral $SU(3)$
Lagrangian predicts the relevance of 12 basically unknown parameters,
$g^{(S)}, g^{(V)}$, $g^{(T)}$ and $Z_{[10]}$, which all need to be adjusted to the
empirical scattering data.
It is important to realize that chiral symmetry is largely predictive in the $SU(3)$ sector
in the  sense that it reduces the number of parameters  beyond
the static $SU(3)$ symmetry. For example one should compare the six tensors which
result from decomposing $8\otimes 8= 1
\oplus 8_S\oplus 8_A \oplus 10\oplus \overline{10}\oplus 27$ into its
irreducible components with the subset of SU(3) structures selected
by chiral symmetry in a given partial wave. Thus static $SU(3)$
symmetry alone would predict 18 independent terms for the s-wave
and two p-wave channels rather than the 11 chiral $Q^2$ background
parameters, $g^{(S)}, g^{(V)}$ and $g^{(T)}$. In our work the number of parameters was
further reduced significantly by insisting on the large $N_c$ sum rules
\begin{eqnarray}
g_1^{(S)}=2\,g_0^{(S)}= 4\,g_D^{(S)}/3 \,, \qquad
g_1^{(V)}=2\,g_0^{(V)}= 4\,g_D^{(V)}/3 \,, \qquad g_1^{(T)}=0 \,,
\nonumber
\end{eqnarray}
for the symmetry conserving quasi-local two body interaction terms (see (\ref{Q^2-large-Nc-result})).
In Tab. \ref{q2param:tab} we collect the values of all free parameters as they result from our
best global fit.
We point out that the large $N_c$ sum rules derived in section 2 implicitly assume that additional inelastic
channels like $K \,\Delta_\mu $ or $K_\mu\,N $ are not too important. The effect of such
channels can be absorbed to some extent into the quasi-local counter terms, however possibly at the
prize that their large $N_c$ sum rules are violated. It is therefore a highly non-trivial result
that we obtain a successful fit imposing (\ref{q2param:tab}). Note that the only previous analysis \cite{Kaiser},
which truncated the interaction kernel to chiral order $Q^2$ but did not include p-waves, found values for the s-wave
range parameters largely inconsistent with the large $N_c$ sum rules. This may be due in part to the use of channel
dependent cutoff parameters and the fact that that analysis missed octet and decuplet exchange contributions,
which are important for the s-wave interaction kernel already to chiral order $Q^2$.

The parameters $b_0, b_D$ and $b_F$ to this order characterize the
explicit chiral symmetry-breaking effects of QCD via the finite current
quark masses.  The parameters $b_D$ and $b_F$ are well estimated
from the baryon octet mass splitting (see (\ref{mass-splitting})) whereas
$b_0$ must be extracted directly from the meson-baryon scattering data.
It drives the size of the pion-nucleon sigma term for which conflicting values are
still being discussed in the literature \cite{pin-news}. Our values
\begin{eqnarray}
b_0 = -0.346\, {\rm GeV}^{-1} \,, \quad  b_D = 0.061\, {\rm GeV}^{-1} \,, \quad  b_F =-0.195 \,{\rm GeV}^{-1} \,,
\label{b-result}
\end{eqnarray}
are rather close to values expected from the baryon octet mass splitting (\ref{mass-splitting}). The
pion-nucleon sigma term $\sigma_{\pi N}$ if evaluated at leading chiral order $Q^2$ would
be $\sigma_{\pi N} \simeq 32$ MeV. That value should not be compared directly with $\sigma_{\pi N}$ as
extracted usually from pion-nucleon scattering data at the Cheng-Dashen point. The required subthreshold
extrapolation involves further poorly convergent expansions \cite{pin-news}. Here we do not attempt to add
anything new to this ongoing debate. We turn to the analogous symmetry-breaking parameters $d_0$ and $d_D$
for the baryon decuplet states. Like for the baryon octet states we use the isospin averaged empirical values
for the baryon masses in any u-channel exchange contribution.
In the s-channel decuplet expressions we use the slightly different values $m^{(\Delta )}_{10} = 1223.2$ MeV and
$m^{(\Sigma )}_{10} = 1374.4$ MeV to compensate for a small mass renormalization induced by the unitarization.
Those values are rather consistent with $d_D \simeq  -0.49$ GeV$^{-1}$ (see (\ref{mass-splitting})).
Moreover note that all values used are quite compatible with the large $N_c$ sum rule
$$b_D+b_F = d_D/3 \,.$$
The parameter $d_0$ is not determined by our analysis. Its determination required the study of the meson baryon-decuplet
scattering processes.

At chiral order $Q^3$ the number of parameters increases
significantly unless additional constraints from QCD are imposed.
Recall for example that \cite{q3-meissner} presents a large
collection of already 102 chiral $Q^3$ interaction terms.
A systematic expansion of the interaction kernel in
powers of $1/N_c$ leads to a much reduced parameter set. For example we established 23 additional
parameters describing the refined 3-point meson-baryon vertices, including in particular explicit
$SU(3)$ symmetry-breaking effects. A detailed analysis of the 3-point vertices in the $1/N_c$ expansion of QCD
reveals that in fact only eleven additional parameters, rather than the 23 parameters, are relevant at leading
order in that expansion. Since the leading parameters $F_R$, $D_R$ together with the symmetry-breaking parameters
describe at the same time the weak decay widths of the baryon octet and decuplet ground states,
the number of free parameters does not increase significantly at the $Q^3$ level if the large $N_c$ limit is applied.
Similarly the $1/N_c$ expansion leads to only four additional parameters describing
the refined two-body interaction vertices. The values of all $Q^3$ parameters can be found in \cite{Lutz:Kolo}.


\subsection{Pion-nucleon scattering}
\renewcommand{\arraystretch}{1.7}
\begin{table}[t]\begin{center}
\begin{tabular}{|c||c|c|c|c|c|}
\hline\hline
 & $\chi $-BS(3) &  KA86\protect\cite{Koch86} &
EM98\protect\cite{EM98} & SP98\protect\cite{SP98} &
GMORW\protect\cite{GMORW}\\ \hline\hline
$a_{[S_-]}^{(\pi N)}$~[fm]& 0.124 & 0.130 &  0.109$\pm$ 0.001 & 0.125$\pm$ 0.001 &
0.116$\pm$ 0.004 \\
$a_{[S_+]}^{(\pi N)}$~[fm]& -0.014 &-0.012 & 0.006$\pm$ 0.001  & 0.000$\pm$ 0.001 & 0.005$\pm$ 0.006\\
$b_{[S_-]}^{(\pi N)}$~[$m_\pi^{-3}]$& -0.007 & 0.008&   0.016 &
0.001$\pm$ 0.001 & -0.009$\pm$ 0.012\\
$b_{[S_+]}^{(\pi N)}$~[$m_\pi^{-3}]$& -0.028 &-0.044&  -0.045 & -0.048$\pm$ 0.001 &
-0.050$\pm$ 0.016\\
\hline\hline
$a_{[P_{11}]}^{(\pi N)}$~[$m_\pi^{-3}$]& -0.083 & -0.078& -0.078$\pm$ 0.003 &
-0.073$\pm$ 0.004 & -0.098$\pm$ 0.005\\
$a_{[P_{31}]}^{(\pi N)}$~[$m_\pi^{-3}$]& -0.045 & -0.044& -0.043$\pm$ 0.002 &
-0.043$\pm$ 0.002 & -0.046$\pm$ 0.004\\
$a_{[P_{13}]}^{(\pi N)}$~[$m_\pi^{-3}$]& -0.038 & -0.030& -0.033$\pm$ 0.003 &
-0.013$\pm$ 0.004 & 0.000$\pm$ 0.004 \\
$a_{[P_{33}]}^{(\pi N)}$~[$m_\pi^{-3}$]& 0.198 & 0.214& 0.214$\pm$ 0.002  & 0.211$\pm$
0.002 & 0.203 $\pm$ 0.002\\ \hline
\end{tabular}
\end{center}
\caption{Pion-nucleon threshold parameters. } \label{pin:th}
\end{table}

We begin with a detailed account of the strangeness zero sector. For a review of pion-nucleon
scattering within the conventional meson exchange picture we refer to \cite{EW}. The various chiral
approaches will be discussed  more explicitly below.
Naively one may want to include the pion-nucleon threshold parameters in a global $SU(3)$ fit. In conventional
chiral perturbation theory the latter are evaluated perturbatively to subleading orders in the chiral
expansion \cite{Bernard,pin-q4}. Here the small pion mass justifies the
perturbative treatment. Unfortunately there is no unique set of threshold
parameters available. This is due to the difficulty in extrapolating the empirical data set
down to threshold, the subtle electromagnetic effects and also some inconsistencies in the data set
itself \cite{pin-news,GMORW}.
A collection of mutual contradicting threshold parameters is collected in Tab.~\ref{pin:th}. In order
to obtain an estimate of systematic errors in the various analyses we confront the
threshold values with the chiral sum rules:
\begin{eqnarray}
&&4\,\pi \left( 1+\frac{m_\pi}{m_N}\right)
a^{(\pi N)}_{[S_-]}=\frac{m_\pi}{2\,f^2}+{\mathcal O}\left(Q^3 \right)\, ,
\nonumber\\
&&4\,\pi \left( 1+\frac{m_\pi}{m_N}\right)
b^{(\pi N)}_{[S_-]}=\frac{1}{4\,f^2\,m_\pi}
-\frac{2\,g_A^2+1}{4\,f^2\,m_N}
+\frac{C^2}{18\,f^2}\,\frac{m_\pi}{m_N\,(\mu_\Delta+m_\pi)}
+{\mathcal O}\left(Q \right)\,,
\nonumber\\
&&4\,\pi \left( 1+\frac{m_\pi}{m_N}\right)
\Big( a^{(\pi N)}_{[P_{13}]}-a^{(\pi N)}_{[P_{31}]} \Big)
=\frac{1}{4\,f^2\,m_N}+{\mathcal O}\left( Q\right)\,,
\nonumber\\
&& 4\,\pi\,\left( 1+\frac{m_\pi}{m_N}\right)
a^{(\pi N)}_{\rm SF}
= -\frac{3\,g_A^2}{2\,f^2\,m_\pi}\left(1+\frac{m_\pi}{m_N}\right)
-\frac23\,\frac{C^2}{f^2}\,\frac{m_\pi\,m_N }
{m_\Delta\,(\mu_\Delta^2-m_\pi^2)}+{\mathcal O}\left( Q\right),
\label{sum-rules}
\end{eqnarray}
where $\mu_\Delta = m_\Delta -m_N $. We confirm the result of \cite{Bernard} that
the spin-flip scattering volume
$a_{\rm SF}^{(\pi N)}=a^{(\pi N)}_{[P_{11}]}+2\,a^{(\pi N)}_{[P_{31}]}-a^{(\pi N)}_{[P_{13}]}
-2\, a^{(\pi N)}_{[P_{33}]}$ and the combination
$a^{(\pi N)}_{[P_{13}]}-a^{(\pi N)}_{[P_{31}]}$ in (\ref{sum-rules})
are independent of the quasi-local 4-point interaction strengths at leading order. Confronting
the analyses in Tab.~\ref{pin:th} with the chiral sum rules quickly reveals that only the EM98
analysis \cite{EM98} appears consistent with the sum rules within 20 $\%$.
The
analysis \cite{GMORW} and \cite{SP98} badly contradict the chiral sum rules (\ref{sum-rules}),
valid at leading chiral orders, and therefore would require unnaturally large correction terms
possibly discrediting the convergence of the chiral expansion in the pion-nucleon sector.
The KA86 analysis is consistent with the two p-wave sum rules but appears inconsistent with
the s-wave range parameter $b_{[S_-]}$. The recent $\pi^-$ hydrogen atom experiment
\cite{Schroeder} gives rather precise values for the $\pi^-$-proton scattering lengths
\begin{eqnarray}
 a_{\pi^-p\to \pi^- p} =&a_{S_-}^{(\pi N)} +a_{S_+}^{(\pi N)}&= \phantom - 0.124\pm 0.001 \;{\rm fm} \;,\qquad
\nonumber\\
 a_{\pi^-p\to \pi^0 n} =& -\sqrt{2}\,a_{S_-}^{(\pi N)} &=-0.180\pm 0.008\; {\rm fm} \;.
\label{pi-atoms}
\end{eqnarray}
These values are in conflict with the s-wave scattering lengths of the EM98 analysis.
For a comprehensive discussion of further constraints from the pion-deuteron scattering
lengths as derived from recent pionic atom data we refer to \cite{gpinn:best}.
All together the emerging picture is complicated and inconclusive at present.
Related arguments are presented by Fettes and Mei\ss ner in their work \cite{pin-q4}
which considers low-energy pion-nucleon phase shifts to chiral order $Q^4$. A resolution for this
puzzle may be offered by the most recent work of Fettes and Mei\ss ner \cite{pin-em} where
they consider electromagnetic corrections term within the $\chi $PT scheme to order $Q^3$.

We turn to a further important aspect to be discussed. Even though the EM98 analysis is
rather consistent with the chiral sum rules (\ref{sum-rules}) does it imply background terms
of natural size?
This can be addressed by considering a further combination of p-wave scattering volumes
\begin{eqnarray}
&&4\,\pi \left( 1+\frac{m_\pi}{m_N}\right)\left( a^{(\pi N)}_{[P_{11}]}
-4\,a^{(\pi N)}_{[P_{31}]}\right) =
\frac{3}{2\,f^2\,m_N} +B+{\mathcal O}\left(Q \right)\;,
\nonumber\\
&&B=-\frac{5}{12 f^2}\,\Big( 2\,\tilde g_0^{(S)}+\tilde g_D^{(S)}+\tilde g_F^{(S)} \Big)
-\frac{1}{3 f^2}\,\Big( \tilde g_D^{(T)}+\tilde g_F^{(T)} \Big) \,,
\end{eqnarray}
where we absorbed the Z-dependence into the tilde couplings.
The naturalness assumption would estimate
$B \sim 1/(f^2\,m_\rho)$, a typical size which is compatible with the background term
$B\simeq 0.92\,m_\pi^{-3}$ of the EM98 solution.

\begin{figure}[h]
\begin{center}
\includegraphics[width=15cm,clip=true]{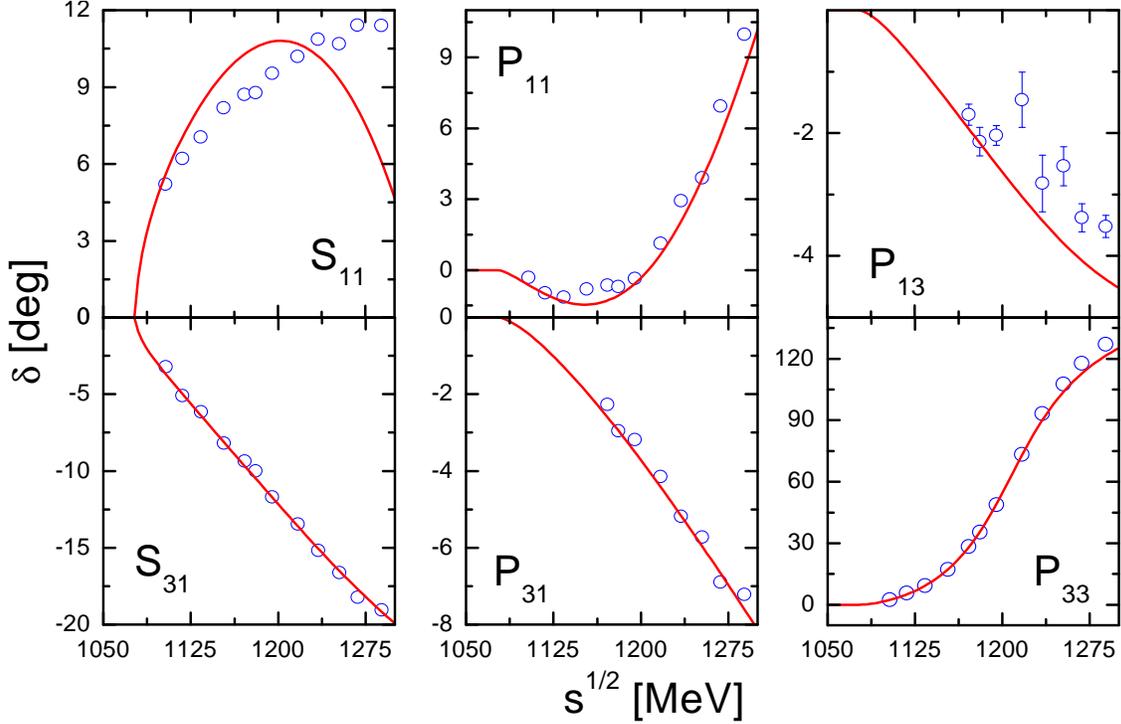}
\end{center}
\caption{S- and p-wave pion-nucleon phase shifts. The single
energy phase shifts are taken from \cite{pion-phases}.}
\label{fig:pionphases}
\end{figure}

In order to avoid the ambiguities of the threshold parameters we decided to
include the single energy pion-nucleon phase shifts of \cite{SP98} in our global fit.
The phase shifts are evaluated in the $\chi$-BS(3) approach including all
channels suggested by the $SU(3)$ flavor symmetry. The single energy phase shifts are fitted
up to $\sqrt{s} \simeq 1200$ MeV.
In Fig.~\ref{fig:pionphases} we confront the result of our fit with the empirical phase shifts.
All s- and p-wave phase shifts are well reproduced up to $\sqrt{s} \simeq 1300$ MeV
with the exception of the $S_{11}$ phase for which our result agrees with the
partial-wave analysis only up to about $\sqrt{s} \simeq 1200$ MeV. We emphasize that one
should not expect quantitative agreement for $\sqrt{s} > m_N+2\,m_\pi \simeq 1215$ MeV where
the inelastic pion production process, not included in this work, starts. The missing higher
order range terms in the $S_{11}$ phase are expected to be induced by additional inelastic
channels or the nucleon resonances $N(1520)$ and $N(1650)$. We confirm the findings of
\cite{Kaiser,new-muenchen} that the coupled $SU(3)$ channels predict considerable strength in the
$S_{11}$ channel around $\sqrt{s} \simeq 1500$ MeV where the phase shift shows a resonance
like structure. Note, however that it is expected that the nucleon resonances $N(1520)$ and
$N(1650)$ couple strongly to each other \cite{Sauerman} and therefore one should not
expect a quantitative description of the $S_{11}$ phase too far away from threshold. Similarly
we observe considerable strength in the $P_{11}$ channel leading to a resonance like structure
around $\sqrt{s} \simeq 1500 $. We interpret this phenomenon as a precursor effect of the p-wave
$N(1440)$ resonance. We stress that our approach differs significantly from the recent work
\cite{new-muenchen} in which the coupled SU(3) channels are applied to pion induced
$\eta$ and kaon production which require much larger energies $\sqrt{s} \simeq m_\eta +m_N \simeq$ 1486 MeV
or $\sqrt{s} \simeq m_K +m_\Sigma \simeq $ 1695 MeV. We believe that such high energies can be
accessed reliable only upon inclusion of additional inelastic channels. The discussion of the
pion-nucleon sector is closed by returning to the threshold parameters. In Tab. \ref{pin:th} our extracted
threshold parameters are presented in the second row. We conclude that all threshold parameters are within
the range suggested by the various analyses.

\subsection{$K^+$-nucleon scattering}

\tabcolsep=1.1mm
\begin{table}[t]\begin{center}
\begin{tabular}{|c|c|c||c|c|c|c|c|}
\hline & $a^{(K N)}_{S_{01}}$ [fm] & $a^{(K N)}_{S_{21}}$ [fm]  &
$a^{(K N)}_{P_{01}}$ $[m_\pi^{-3}]$ & $a^{(K N)}_{P_{21}}$
$[m_\pi^{-3}]$ & $a^{(K N)}_{P_{03}}$ $[m_\pi^{-3}]$ & $a^{(K
N)}_{P_{23}}$ $[m_\pi^{-3}]$  \\ \hline \hline $\chi$-BS(3)  &
0.06 & -0.30 & 0.033 & -0.017 & -0.003 & 0.012 \\ \hline
\protect\cite{Hyslop} & 0.0 & -0.33 & 0.028 & -0.056 & -0.046 &
0.025 \\ \hline \protect\cite{BR:Martin} & -0.04 & -0.32 & 0.030 &
-0.011 & -0.007 & 0.007 \\ \hline
\end{tabular}
\end{center}
\caption{$K^+$-nucleon threshold parameters. The values of the
$\chi$-BS(3) analysis are given in the first row. The last two
rows recall the threshold parameters as given in \cite{Hyslop} and
\cite{BR:Martin}.} \label{tab-r-kp}
\end{table}

We turn to the strangeness plus one channel. Since it is impossible to give here a comprehensive
discussion of the many works dealing with kaon-nucleon scattering we refer to the review
article by Dover and Walker \cite{Dover} which is still up to date in many respects.
To summarize the data situation: there exist precise low-energy differential cross
sections for $K^+ p$ scattering but no scattering data for the $K^+$-deuteron scattering
process at low energies. Thus all low-energy results in the isospin zero
channel necessarily follow from model dependent extrapolations.
In our global fit we include the available differential cross section.

At this place it is instructive to consider the threshold amplitudes in some details.
In the $\chi$-BS(3) approach
the s-wave scattering lengths are renormalized strongly by unitarization. At leading order the s-wave
scattering lengths are
\begin{eqnarray}
4\,\pi\left( 1+\frac{m_K}{m_N}\right) a^{(KN)}_{S_{21}} &=&
-m_K \left( f^2+\frac{m^2_K}{8\,\pi}
\left(1-\frac{1}{\pi}\,\ln \frac{m_K^2}{m_N^2} \right)\right)^{-1} \;,
\nonumber\\
4\,\pi\left( 1+\frac{m_K}{m_N}\right) a^{(KN)}_{S_{01}} &=&0 \;.
\end{eqnarray}
This leads to $a^{(KN)}_{S21} \simeq -0.22$ fm and $a^{(KN)}_{S01} =0$ fm
close to our final values at subleading orders as given in Tab. \ref{tab-r-kp}.
In that table we collected  typical results for the p-wave scattering volumes also.
The large differences in the isospin zero channel reflect the fact that
that channel is not constraint by scattering data directly \cite{Dover}.
We find that our p-wave scattering volumes, also shown in Tab. \ref{tab-r-kp}, differ in part
significantly from the values given by previous analyses. Such discrepancies may be explained by
important cancellation mechanisms among the u-channel baryon octet and decuplet contributions.
An accurate description of the scattering volumes requires a precise input for the meson-baryon
3-point vertices. Since the $\chi$-BS(3) approach describe the 3-point vertices in accordance
with all chiral constraints and large $N_c$ sum rule of QCD we believe our values for
the scattering volumes to be quite reliable.

\begin{figure}[h]
\begin{center}
\includegraphics[width=15.5cm,clip=true]{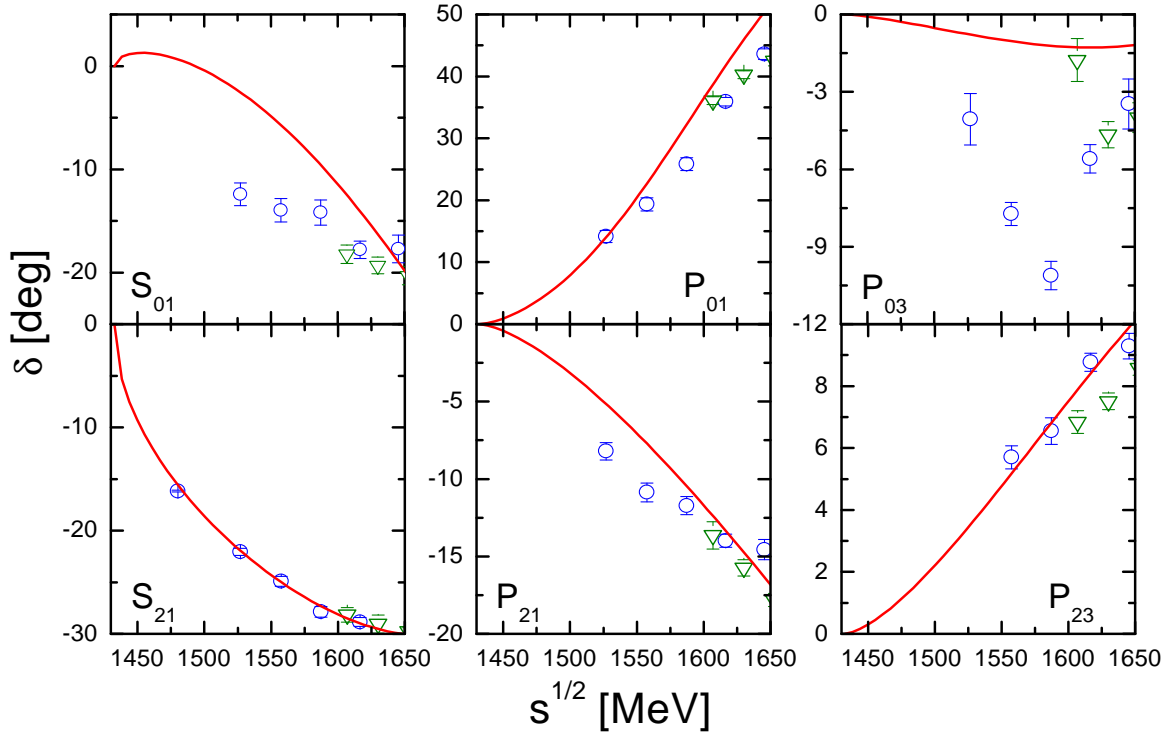}
\end{center}
\caption{S- and p-wave $K^+$-nucleon phase shifts. The solid lines
represent the results of the $\chi$-BS(3) approach. The open
circles are from the Hyslop analysis \cite{Hyslop} and the open
triangles from the Hashimoto analysis \cite{Hashimoto}.}
\label{fig:K+phases}
\end{figure}

In Fig.~\ref{fig:K+phases} we confront our s- and p-wave $K^+$-nucleon
phase shifts with the most recent analyses by Hyslop et al.~\cite{Hyslop}
and Hashimoto~\cite{Hashimoto}.
We find that our partial wave shifts are reasonably close to the single energy phase shifts of
\cite{Hyslop} and \cite{Hashimoto} except the $P_{03}$ phase for which we obtain much smaller strength.
Note however, that at higher energies we smoothly reach the single energy phase shifts of Hashimoto \cite{Hashimoto}.
A possible ambiguity in that phase shift is already suggested by the conflicting scattering volumes found in that
channel in earlier works (see Tab. \ref{tab-r-kp}). The isospin one channel, on the other hand, seems well established
even though the data set does not include polarization measurements close to threshold which are needed
to unambiguously determine the p-wave scattering volumes.

\subsection{$K^-$-nucleon scattering}

It is left to report on our results in the strangeness minus one sector. The antikaon-nucleon scattering
process shows a large variety of intriguing phenomena. Inelastic channels are already open at threshold
leading to a rich coupled-channel dynamics. Also the $\bar K N$ state couples to many of the observed
hyperon resonances for which competing dynamical scenarios are conceivable.
We fit directly to the available data set rather than to any partial wave
analysis. Comparing for instance the energy dependent analyses \cite{gopal} and \cite{garnjost}
one finds large uncertainties in the s- and p-waves in particular at low energies. This reflects on the
one hand a model dependence of the analysis and on the other hand an insufficient data set. A partial wave
analysis of elastic and inelastic antikaon-nucleon scattering data without further constraints from theory is
inconclusive at present \cite{Dover,Gensini}. For a detailed overview of the more ancient theoretical
analyses we suggest the review article by Dover and Walker \cite{Dover}.

\begin{figure}[t]
\begin{center}
\includegraphics[width=15.5cm,clip=true]{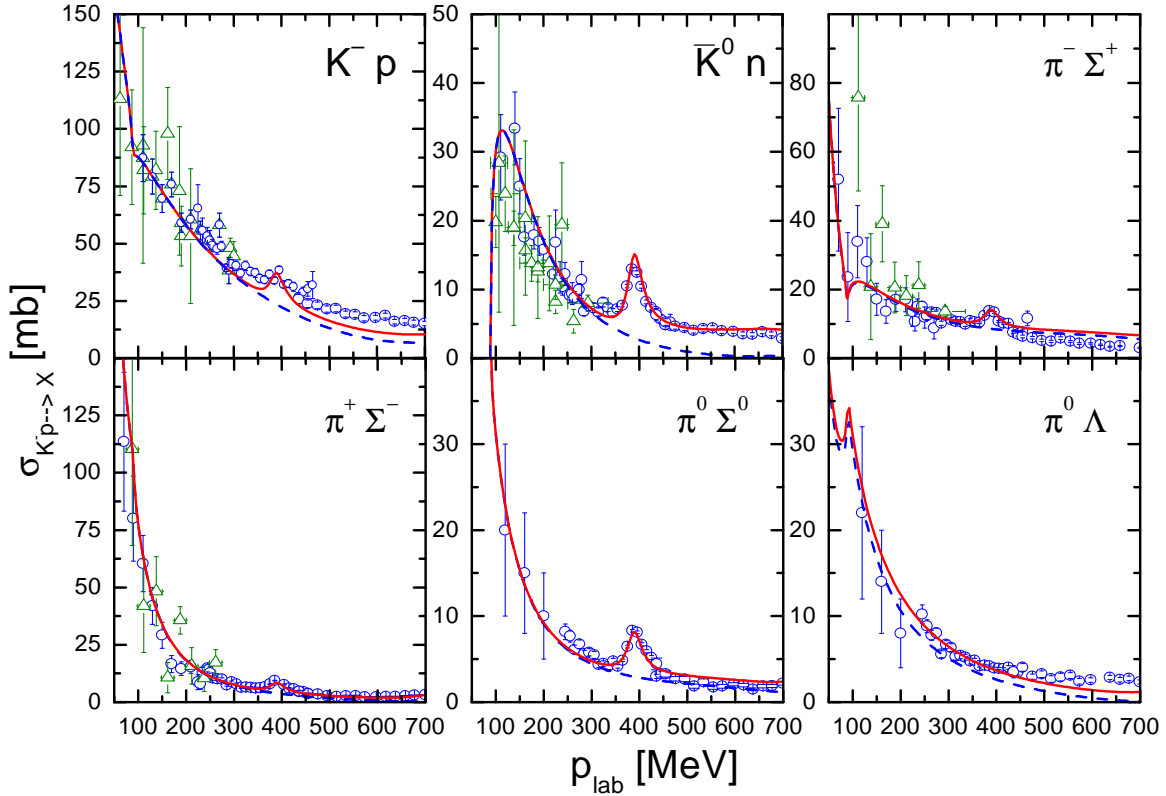}
\end{center}
\caption{$K^-$-proton elastic and inelastic cross sections. The
data are taken from \cite{ref4a,ref4circles,old-scat}. The solid
lines show the results of our $\chi$-BS(3) theory including all
effects of s-, p- and d-waves. The dashed lines represent the
s-wave contributions only. We fitted the data points given by open
circles~\cite{ref4a,ref4circles}. Further data points represented
by open triangles~\cite{old-scat} were not considered in the
global fit.} \label{fig:totcross}
\end{figure}

In Fig.~\ref{fig:totcross} we present the result of our fit for the
elastic and inelastic $K^-p$ cross sections. The data set is nicely reproduced including
the rather precise data points for laboratory momenta 250 MeV$<p_{\rm lab}<$ 500 MeV.
In Fig.~\ref{fig:totcross} the s-wave contribution to the total cross section is shown with a dashed line.
Sizeable p-wave contributions are found at low energies only in the $\Lambda \pi^0$ production cross section.
The $\Lambda \pi^0$ channel carries isospin one and therefore provides a valuable
constraint on the poorly known $K^-$-neutron interaction.
Note that according to~\cite{watson} the inelastic channel $K^-p \to \Lambda \pi \pi$,
not included in this work, is no longer negligible at a quantitative level
for $p_{\rm lab} > 300$~MeV.

Important information on the p-wave dynamics is provided by angular
distributions for
the inelastic $K^-p$ reactions. {The available data are represented in terms of
coefficients $A_n$  characterizing the differential cross section
$d\sigma(\cos \theta , \sqrt{s}\,) $  as a function of the center of mass
scattering angle $\theta $ and the total energy $\sqrt{s}$}
\begin{eqnarray}
\frac{d\sigma (\sqrt{s}, \cos \theta )}{d\cos \theta }  &=&
\sum_{n=0}^\infty A_n(\sqrt{s}\,)\,P_n(\cos \theta )  \;,
\label{a-b-def}
\end{eqnarray}
where $P_n(\cos\theta)$ are the Legendre polynomials.
\begin{figure}[t]
\begin{center}
\includegraphics[width=14cm,clip=true]{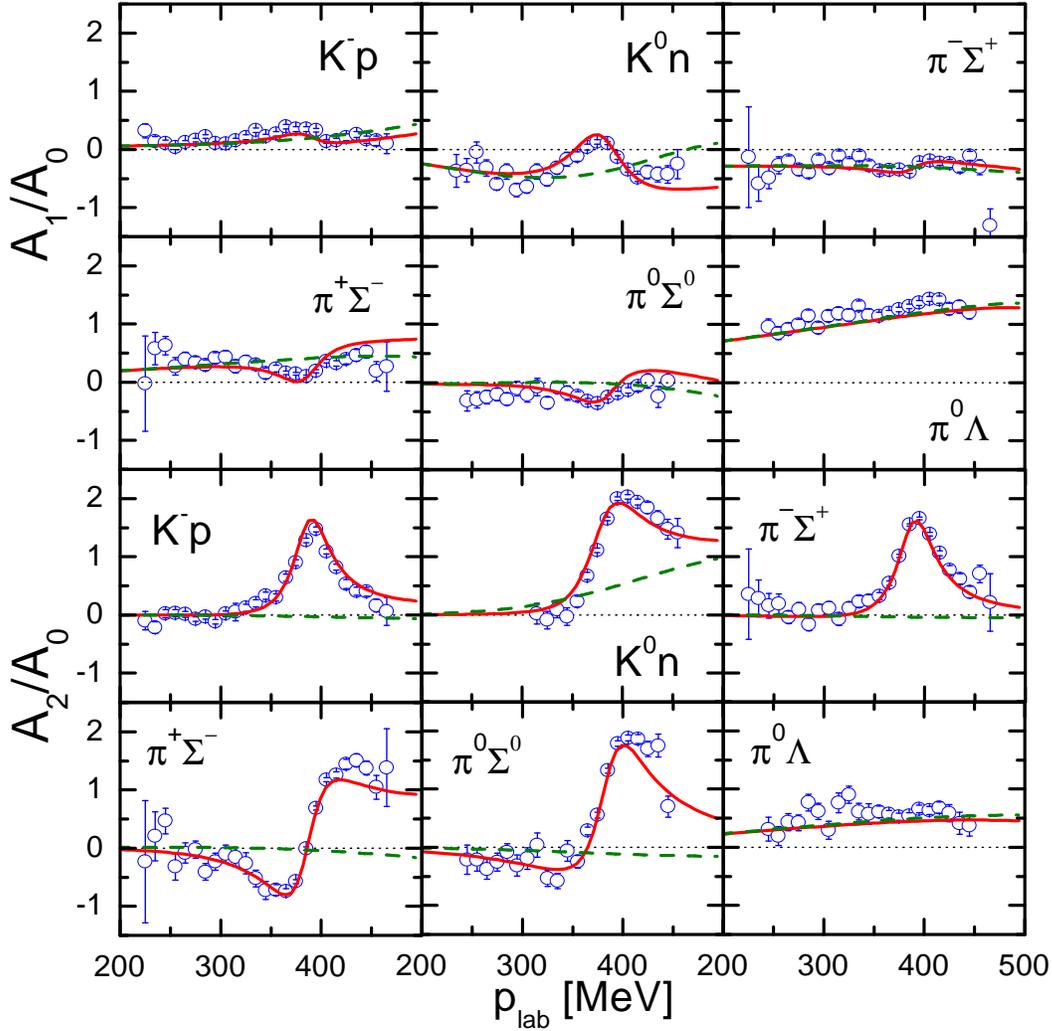}
\end{center}
\caption{Coefficients $A_1$ and $A_2$ for the $K^-p\to \pi^0 \Lambda$,
$K^-p\to \pi^\mp \Sigma^\pm$
and $K^-p\to \pi^0 \Sigma$ differential cross sections. The data are
taken from \cite{ref4a}.}
\label{fig:a}
\end{figure}
In Fig.~\ref{fig:a} we compare the empirical ratios $A_1/A_0$ and $A_2/A_0$ with the results of
the $\chi$-BS(3) approach. Note that for $p_{\rm lab} < 300$ MeV the empirical ratios with $n\geq 3$ are
compatible with zero within their given errors. A large $A_1/A_0$ ratio
is found only in the $K^-p\to \pi^0 \Lambda$ channel demonstrating again the importance of
p-wave effects in the isospin one channel. The dotted lines of Fig.~\ref{fig:a} confirm the importance
the $\Lambda(1520)$ resonance for the angular distributions in the isospin zero channel. The fact that
this resonance appears more important in the differential cross sections than in the total cross sections
follows simply because the tail of the resonance is enhanced if probed via an interference term. In the
differential cross section the $\Lambda(1520)$ propagator enters linearly whereas the total cross section
probes the squared propagator only.

\section{Summary}

We summarize the main issues reported on in this review. The recently developed
$\chi$-BS(3) approach was presented in some detail demonstrating its successful
application to meson-baryon scattering. In that approach the Bethe-Salpeter interaction kernel
was evaluated to chiral order $Q^3$ where the number of parameters was reduced significantly
by performing a systematic $1/N_c$ expansion of the interaction kernel. The scattering amplitudes,
used to obtain the differential cross sections, are the solution of the covariant Bethe-Salpeter
scattering equation. The first reliable estimates of previously poorly known s- and p-wave parameters of the
chiral SU(3) Lagrangian were obtained. It is a highly non-trivial and novel result that the
strength of all quasi-local 2-body interaction terms are consistent with the expectation from the
large $N_c$ sum rules of QCD. Moreover, all parameters established prove the chiral $SU(3)$ flavor symmetry
to be an extremely useful and accurate tool. Explicit symmetry breaking effects are quantitatively important
but sufficiently small to permit an evaluation within our $\chi$-BS(3) approach. This confirms a beautiful analysis
by Hamilton and Oades \cite{Hamilton:Oades} who strongly supported the $SU(3)$ flavor symmetry
by a discrepancy analysis of kaon-nucleon scattering data. The $\chi$-BS(3) approach establishes a unified
description of pion-nucleon and kaon-nucleon scattering describing a large amount of empirical scattering
data including the axial-vector coupling constants for the baryon octet ground states.
An important test of the $\chi$-BS(3) analysis could be provided by new data on kaon-nucleon scattering
from the DA$\Phi$NE facility \cite{DAPHNE}.


\end{document}